\documentclass[prx,aps,amssymb,amsmath,reprint,superscriptaddress,showpacs,floatfix,longbibliography]{revtex4-1}

\usepackage{bm}
\usepackage{graphicx}
\usepackage{color}
\usepackage[b]{esvect} % Vector notation 
\usepackage{mathtools}
\usepackage{braket}
\usepackage{amsfonts}
\usepackage{textcomp} % text companion fonts
\usepackage{microtype} 
\usepackage[normalem]{ulem} %to strike the words
\usepackage{hyperref}

\hypersetup{colorlinks=true}
\usepackage[all]{hypcap} % let hyperlinks correctly point to figures rather than their captions; 
\newcommand{\akd}{a^{\dagger}_{k}}
\newcommand{\ak}{a^{\phantom{e\dagger}}_{k}}
\newcommand{\aked}{a^{e\dagger}_{k}}
\newcommand{\ake}{a^{e\phantom{\dagger}}_{k}}
\newcommand{\akod}{a^{o\dagger}_{k}}
\newcommand{\ako}{a^{o\phantom{\dagger}}_{k}}
\newcommand{\w}{\omega}

\newcommand{\da}{\downarrow}
\newcommand{\mr}[1]{\mathrm{#1}}

\newcommand{\eq}[1]{\begin{align}#1\end{align}}
\newcommand{\ba}{\begin{array}}
\newcommand{\ea}{\end{array}}
\newcommand{\bit}{\begin{itemize}}
\newcommand{\eit}{\end{itemize}}

\newcommand{\nn}{\nonumber}

\newcommand{\f}{\frac}

\newcommand{\mrm} {\mathrm}

\newcommand{\ra}{\rangle}
\newcommand{\la}{\langle}
\newcommand{\ua}{\uparrow}

\newcommand*\conj[1]{{#1^*}}

\begin{document}

%\title{Non-linear scattering of coherent states in ultra-strongly coupled waveguides}
%\title{Particle Production in many-body quantum circuits}
\title{Particle Production in Ultra-Strong Coupling Waveguide QED}

\author{Nicolas Gheeraert}
\affiliation{Institut N\'{e}el, CNRS and Universit\'e Grenoble Alpes, F-38042 Grenoble, France}
\author{Xin H. H. Zhang}
\affiliation{Department of Physics, Duke University, P. O. Box 90305,
Durham, North Carolina 27708, USA}
\author{Th\'eo S\'epulcre}
\affiliation{Institut N\'{e}el, CNRS and Universit\'e Grenoble Alpes, F-38042 Grenoble, France}
\author{Soumya Bera}
\affiliation{Department of Physics, Indian Institute of Technology Bombay,
Mumbai 400076, India}
\author{Nicolas Roch}
\affiliation{Institut N\'{e}el, CNRS and Universit\'e Grenoble Alpes, F-38042 Grenoble, France}
\author{Harold U. Baranger}
\affiliation{Department of Physics, Duke University, P. O. Box 90305,
Durham, North Carolina 27708, USA}
\author{Serge Florens}
\affiliation{Institut N\'{e}el, CNRS and Universit\'e Grenoble Alpes, F-38042 Grenoble, France}

\begin{abstract}
Understanding large-scale interacting quantum matter requires dealing with the
huge number of quanta that are produced by scattering even a few particles
against a complex quantum object. Prominent examples are found from high energy
cosmic ray showers, to the optical or electrical driving of degenerate Fermi
gases. We tackle this challenge in the context of many-body quantum optics, as
motivated by the recent developments of circuit quantum electrodynamics at
ultrastrong coupling. The issue of particle production is addressed quantitatively 
with a simple yet powerful concept rooted in the quantum superposition principle 
of multimode coherent states. This key idea is illustrated by the study of 
multi-photon emission from a single two-level artificial atom coupled to a high
impedance waveguide, driven by a nearly-monochromatic coherent tone. We find 
surprisingly that the off-resonant inelastic emission lineshape is dominated by 
broadband particle production, due to the large phase space associated with 
contributions that do not conserve the number of excitations. Such frequency 
conversion processes produce striking signatures in time correlation measurements, 
which can be tested experimentally in quantum waveguides. These ideas open new 
directions for the simulation of a variety of physical systems, from polaron 
dynamics in solids to complex superconducting quantum architectures.
\end{abstract}

\date{\today}

\maketitle

\section{Introduction}

Exploring the quantum world~\cite{HarocheRaimondBook} is an ongoing quest
fueled by the search for fundamental understanding, which has
enabled the creation of unexpected technologies.
The advent of lasers and semiconducting microelectronics has indeed crucially
relied on building blocks that are determined at the microscopic level by
quantum effects. Whether intrinsically quantum effects such as entanglement can
provide further practical scientific developments is at present intensely
investigated. However, addressing increasingly complex quantum systems is
pushing the boundaries of what simulations can cope with on present day
hardware, due to the exponential complexity growth when working with states from
the Hilbert space. This question is certainly very acute when dealing with the
temporal driving of large-scale quantum circuits, which can lead to a rapid
proliferation of propagating quanta. How to encode quantum information efficiently in such a
situation, using only available classical computers, is a very general challenge
in contemporary physics.

Because quantum many-body scattering is relevant for a wide range of physical systems 
(solid state materials, cold atomic gases, high energy collisions in particle
accelerators), fruitful concepts are best developed with the relevant physics at
hand. For this reason, we focus in this article on the topic of many-body
quantum optics, which combines discrete atomic states (the scatterer) with
broadband photonic fields (leading to a huge Hilbert space of quanta).
Historically, light-matter interaction has been thoroughly studied in the regime of standard 
quantum optics~\cite{meystre_elements_2010,LoudonQTL03}, where the combination of small 
atomic dipoles and perturbative fine structure constant $\alpha_\mr{QED}\simeq1/137$ 
leads to small radiative corrections, such as the famous Lamb shift at order 
$[\alpha_\mr{QED}]^3$ (in units of the atomic frequencies).
Quantum electrodynamics (QED) corrections to the bare atom picture also control the natural 
linewidth of atomic transitions~\cite{LoudonQTL03,eikema_linewidth_2001} associated to vacuum fluctuations of
the electromagnetic field, occurring also at third order in $\alpha_\mr{QED}$. 
As a consequence, the electromagnetic modes that may strongly interact with an atom 
are limited to those very close to its resonance frequency. A variety of strategies are
being pursued in atomic quantum optics in order to enhance the strength of light-matter 
coupling. First, there is extensive work on confining light to a cavity in order to 
%decrease its mode volume and so 
increase the magnitude of the electric field
\cite{MabuchiSci02,HarocheRaimondBook,HarocheRMP2013}; however, in this case,
interesting effects involving a photon continuum are discarded. Under strong
pumping, multi-photon non-resonant contributions can become sizable, but this
suffers from the same problem of rather limited bandwidth.
Finally, several strategies involving photonic crystals or Rydberg
atoms are being pursued in which a collective light-matter coupling is made
strong by using a large number of weakly coupled components
\cite{HartmannQsimJOPT16,HaapamakiBajcsy16,SolanoNanofibersAAMOP17,RoyRMP17}. 

We wish to address, however, regimes where radiative effects become of order one 
in a system with a \textit{single} emitter and a \textit{broad continuum} of 
photonic modes, an area known as ultra-strong coupling waveguide quantum
electrodynamics (wQED). Access to this regime is becoming
possible~\cite{forndiaz_galvanic_2017,puertas_transmon_2017} through circuit
quantum electrodynamics in which artificial superconducting atoms interact
on-chip with microwave transmission lines (see Refs.~\cite{RoyRMP17,nori_review_2017} 
for general reviews on the topic); in fact, in cavities, ultra-strong coupling has
been achieved in this
system~\cite{devoret_cavity_2007,schoelkopf_cavity_2008,bourassa_cavity_2009,
niemczyk_cavity_2010,forndiaz_cavity_2010,yoshihara_DSC_2017}.  In ultra-strong
wQED, many-body phenomena are expected to occur that have no counterpart in
standard quantum optics~\cite{meystre_elements_2010,LoudonQTL03} or in
low-coupling superconducting transmission
lines~\cite{astafiev_fluorescence_2010,abdumalikov_emission_2011,
hoi_router_2011,hoi_g2,hoi_mollow,vanloo_interaction_2013,sundaresan_multimode_2015}.
A non-exhaustive list of theoretical predictions includes giant Lamb
shifts~\cite{leggett_dynamics_1987,lehur_kondo_2012,peropadre_2013,bera_stabilizing_2014,diaz_polaron_2016},
single-photon
down-conversion~\cite{goldstein_inelastic_2013,sanchez_inelastic_2014}, non-RWA
transmission
lineshapes~\cite{lehur_kondo_2012,bera_dynamics_2016,shi_scattering}, multi-mode
entanglement~\cite{bera_generalized_2014,snyman_robust_2015,shi_squeezed}, and
non-classical emission~\cite{gheeraert_spontaneous_2016}.  The key element in
all of the novel many-body phenomena in ultra-strong wQED is that the number of
excitations is no longer conserved because the rotating-wave approximation is
not legitimate anymore. It is worthwhile then to focus directly on this non-conservation.  We
show here that a key signature of scattering or excitation in the ultra-strong
regime is broadband photon production: a greater number of photons come out than
go in, even in the very low power single-photon excitation regime. 

In contrast to previous studies which focused on effects that become prominent when 
the light-matter coupling $\alpha$ reaches values of order one (the so-called Kondo regime), 
we investigate here many-body effects that are realistically 
observable when entering the ultra-strong coupling regime, with typically 
$0.1\lesssim \alpha\lesssim0.3$. These many-body effects are nevertheless dramatic and have the additional advantage that they may be probed experimentally in the very near future. 
This regime is characterized by a qubit linewidth $\Gamma$ that is a sizeable fraction of 
its resonance frequency $\Delta$, owing to the perturbative relation 
$\Gamma\simeq\pi \alpha \Delta$. For $\alpha\ll1$, it is widely believed in 
the quantum optics context that dominant physical processes are well captured by the so-called 
rotating wave approximation (RWA), upon which non-resonant transitions are discarded from the outset. 
While it is true that RWA provides quantitatively accurate results for the linear response of an 
atom weakly coupled to a waveguide, we find that low-power non-linear scattering properties are 
however dominated by non-RWA contributions, even for arbitrarily small coupling
in which the RWA is thought to become exact.

\begin{figure}[t]
\includegraphics[width=1.0\linewidth]{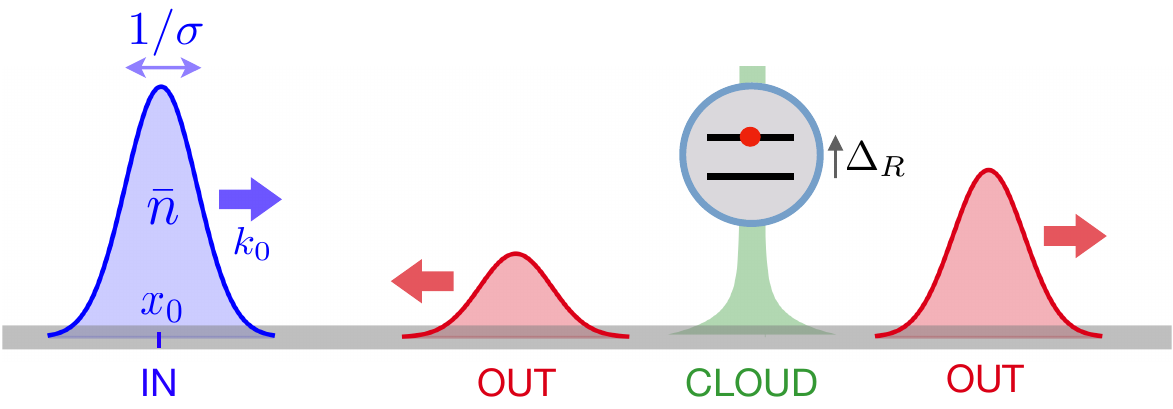}
\caption{Illustration of the setup considered in this paper: a long transmission
waveguide (grey horizontal line) is side-coupled to a two-level system, allowing 
the measurement of multi-photon scattering matrices in a typical two-terminal 
geometry from the reflection and transmission (outgoing states with arrows
pointing outward) of a coherent 
state Gaussian wavepacket (incoming state with arrow pointing inward). A many-body polarization cloud 
lives in the central region (tied to the qubit)~\cite{snyman_robust_2015}. 
Frequency-conversion processes are extracted by a spectral and number state analysis 
of the outgoing wavepackets.}
\label{waveguide}
\end{figure}

The scenario that we consider is shown in Fig.~\ref{waveguide}. A right-going 
coherent state pulse is injected into a waveguide. The waveguide and qubit are 
initially in their ground state, implying that the qubit is non-perturbatively dressed by 
a cloud of waveguide photons~\cite{snyman_robust_2015}. The incoming coherent
state pulse then scatters from this dressed state, leading to outgoing transmitted 
and reflected pulses, that have acquired on general grounds a many-body
character~\cite{shi_Smatrix}.

Our goal here is two-fold. First, we uncover new physical effects in non-linear many-body photon scattering
%, and to guide its experimental measurement 
by analyzing the photonic
content of non-resonant emission spectra. One major observation is that
significant non-linear emission arises from both RWA and non-RWA pathways. 
In light of standard knowledge in quantum optics,
it comes as a surprise that non-RWA processes are found to dominate in magnitude the 
RWA non-linear response when off resonance. 
%, even for processes that emerge from the same input
%states (and thus scale {\it a priori} identically as a function of 
%the incoming beam power). 
%This effect may be interpreted as a
%consequence of the larger phase space of outgoing states for particle
%production, which is intrinsically broadband.
Indeed, in the regime of ultra-strong coupling, the linewidth of the qubit broadens
substantially, leading to important non-resonant inelastic contributions to the scattering 
cross-section. Under a drive that is detuned in frequency above the resonance 
of the qubit, for instance, inelastic down-conversion occurs by the ``splitting"
of an incoming  photon into several lower energy
ones~\cite{goldstein_inelastic_2013}. For larger power, there are similar
processes involving an increasing number of incoming photons, all of which are
described by counter-rotating terms. These processes are favored by a wide
continuum of available outgoing multi-photon states.  The surprising dominance
of non-RWA processes can thus be interpreted as a consequence of the larger
phase space of outgoing states for particle production. 
%This process, among others 
%involving an increasing number of incoming photons for larger drive power and that are 
%described by counter-rotating terms, is favored by a wide continuum of available outgoing multi-photon states. 

Another dramatic many-body effect is uncovered by studying the correlations in
time. We find that ultra-strong coupling leads to striking qualitative
signatures in the photon statistics of a single emitter, namely incomplete
anti-bunching on resonance at zero time-delay and strong bunching at finite
delay, that are very prominent in the off-resonant case.  
These previously unrecognized features are quantitatively different from RWA 
results and constitute important signatures of particle production from an 
experimental point of view.  A further indication of interesting many-body effects 
is that perturbative expansions for the elastic and inelastic emission spectra 
cannot be captured quantitatively.

Our second objective is to provide a general and powerful simulation toolbox 
to access the non-linear and inelastic processes involved to any
order in the incoming beam power. This methodology relies on an expansion of
the full many-body wavefunction (qubit+waveguide) in terms of multi-mode coherent states,
using quantum superpositions of several classical-like configurations. It was 
introduced recently as a numerically controlled technique to capture the ground 
state~\cite{bera_stabilizing_2014,bera_generalized_2014} or quenched 
dynamics~\cite{gheeraert_spontaneous_2016} in ultra-strong coupling wQED. 
Two original developments are made in the present manuscript. First, a new
and more numerically efficient algorithm is proposed, which allows for the first
time to tackle in a controlled way the many-body dynamics in waveguides composed of 
several thousands of modes. 
Second, we develop a many-body scattering protocol, that can be used to
simulate realistic scattering setups, as shown in
Fig.~\ref{waveguide}, allowing to deal with the challenging problem of many-body 
particle production in quantum optics.
%, whereupon coherent states wavepackets 
%are impinging on a non-perturbatively dressed atom, 
The resulting multi-photon emission processes in the output field 
are characterized precisely. Despite the coupling being
weak to intermediate in magnitude, non-RWA contributions to these
multi-photon processes open the door to a tremendously large Hilbert space. 
Typically our calculations manage up to five photons 
in the outgoing beam, which, for a long waveguide accounting for about 1000 environmental modes, 
leads to an effective Hilbert space of order $10^{15}\simeq2^{50}$. It is quite remarkable that a quantum 
superposition of classical-like multi-mode coherent states can be harnessed as an efficient 
computing resource to address quantum many-body problems that are currently well beyond the 
reach of any brute force numerical method.

Regarding waveguides, several strategies are possible in order to bring these systems
into a truly many-body territory, such as using the inductive coupling of a flux qubit 
to a low-impedance coplanar waveguide transmission line~\cite{forndiaz_galvanic_2017,haeberlein_spin-boson_2015}, 
or tayloring a charge-like qubit with a capacitive coupling to a high impedance 
metamaterial~\cite{lehur_kondo_2012,goldstein_inelastic_2013,snyman_robust_2015}. 
For this latter purpose, long chains of Josephson 
junctions~\cite{masluk_microwave_2012,bell_quantum_2012,altimiras_tunable_2013,
weissl_bloch_2015} constitute a promising platform that is currently under
investigation~\cite{puertas_transmon_2017} in the context of multi-mode
ultra-strong coupling quantum optics.
In any case, it remains challenging at present to control experimentally a strongly non-linear 
element constituting a true two-level system (such as the Cooper pair box or a flux qubit) 
that is also very well coupled to a designed environment, because non-linearity
brings a high sensitivity to external noise sources. Designs based on a weakly 
non-linear qubit, such as a transmon~\cite{koch_transmon_2007} ultra-strongly coupled to a
waveguide~\cite{blais_transmonUSC_2016,puertas_transmon_2017} could offer an interesting 
alternative for high precision measurements, at the expense however of weakening the 
sought-after non-linear effects.
 
The paper is organized as follows. We first review in Sec.~\ref{Sec2} the basic
model of waveguide quantum electrodynamics, and
develop a general many-body wavefunction approach for the study of inelastic
photon emission by a single two-level system. Section~\ref{Sec3} presents
detailed inelastic emission spectra, in connection with the relevant physical processes.
Section~\ref{sec4} provides a comparison to standard results in quantum optics,
based on the RWA, which can only account for processes in which two input photons
are inelastically scattered, keeping the number of outgoing photons equal to
two. This section closes with a discussion of the temporal correlations of
the emitted light, showing several qualitative features of ultrastrong coupling.
Finally, the perspectives section, Sec.~\ref{Sec5},
discusses prospects for experimental measurements of these effects in
superconducting circuits, and the need for developing further our theoretical
tools in order to capture realistic aspects of Josephson waveguides beyond the
spin-boson limit. Appendices contain technical derivations that should make the
manuscript self-contained, and present details on the new algorithm proposed in
this work.

\section{Many-body coherent state scattering formalism}
\label{Sec2}

\subsection{Modeling a two-level system coupled to a waveguide}

The main assumption that will be made in this study is the restriction of the
atom to a perfect two-level system. This hypothesis is perfectly legitimate 
for strongly non-linear qubits, such as the Cooper pair box or the flux 
qubit~\cite{astafiev_fluorescence_2010,AbdumalikovPRL10,forndiaz_galvanic_2017,NIBA_2017}, 
although these devices typically experience more strongly charge or flux noise
compared to a transmon qubit (which is however weakly non-linear).
Focusing on a two-level system aims to capture the maximum inelastic scattering 
cross-sections, due to its intrinsically high non-linearity. It is thus an 
excellent testbed to examine physics that is already quite rich, and to develop 
state-of-the-art methodologies in the most challenging situation 
from a computational point of view. Following this path, a qubit coupled to a full
one-dimensional waveguide is quite generically expressed by the so-called spin-boson 
Hamiltonian (setting $\hbar$ to unity):
\begin{equation}
H = \frac{\Delta}{2} \sigma_x - \frac{\sigma_z}{2}\sum_{k \in \mathbb{R}} 
g_k^\mr{full} (\ak + \akd) + \sum_{k\in\mathbb{R}} \w_k \akd \ak,
\label{SBinit}
\end{equation}
with $\Delta$ the bare splitting of the qubit levels. We stress that we
do not work in the qubit eigenbasis here, but rather in a basis that makes
the qubit-waveguide coupling diagonal, as described by the $\sigma_z$
term above (this corresponds for instance to the charge basis for a Cooper pair 
box that is capacitively coupled to a waveguide). This choice allows a natural 
description of the driving force behind the entanglement between the qubit and 
the waveguide, and sets the natural language for our numerical technique based on 
coherent states. The momentum dependence of the coupling 
constant $g_k^\mr{full}$ to mode $\akd$ of the full waveguide depends on the device geometry 
and its physical parameters, such as inter-island capacitances, ground capacitances 
and inter-island Josephson energy. In the case where the waveguide is constructed 
from a Josephson junction array,
Refs.~\cite{goldstein_inelastic_2013,snyman_robust_2015,parra_2017} 
proposed explicit microscopic derivations of the coupling constants based on rather 
different designs. Similarly, the momentum dispersion of the eigenfrequencies $\w_k$ 
of the photonic modes is determined by the microscopic details of the waveguide.

In what follows we will consider for simplicity a linear dispersion relation given 
by $\omega_k=|k|$ (taking the speed of light in the metamaterial $c=1$), and a
simple parametrization of the coupling constant.
For this purpose, and in order to simplify the problem, we start by folding the 
bosonic modes of the full waveguide onto a half-line, by defining even and odd 
modes:
\begin{equation}
a_{k}^{e}=\frac{1}{\sqrt{2}}\left(a_{k}+a_{-k}\right)\quad\textrm{and}\quad
a_{k}^{o}=\frac{1}{\sqrt{2}}\left(a_{k}-a_{-k}\right),
\end{equation}
so that the Hamiltonian~(\ref{SBinit}) can be rewritten as
\begin{equation}
H = \frac{\Delta}{2} \sigma_x - \frac{\sigma_z}{2}\sum_{k>0}
g_k (\ake + \aked) + \sum_{k>0} \w_k [\aked \ake\!+\akod \ako],
\label{SB}
\end{equation}
with the coupling constant to the even modes, $g_k = \sqrt{2}
g_k^\mr{full}$.
We choose a parametrization of the effective coupling constant $g_k$ given by 
the following spectral function:
\begin{equation}
J(\w) = \sum_{k>0} \pi g_k^2 \delta(\w-\w_k)
= 2 \pi \alpha\, \w\, \mrm{e}^{-\omega/\omega_c}.
\label{SpectralDensity}
\end{equation}
This form of spectral function, although not completely generic, contains the 
main realistic ingredients of the qubit-waveguide interaction, such as a linear 
ohmic frequency dependence at low energy, and a rapid falloff near the plasma edge
$\omega_c$, that we assume to be exponential in form. For a discretized momentum
grid, we deduce that the coupling constant $g_k$ to even modes reads:
\eq{
\label{gk}
g_k = \sqrt{2\, \alpha \, \omega_k\, \delta k \,
\mrm{e}^{-\omega_k/\omega_c}},
}
where $\delta k$ is the wave-number spacing corresponding to the discretisation
of the continuous momentum integral.

In the form of Hamiltonian~(\ref{SB}), only the even modes are interacting with the 
qubit, while the odd modes are freely propagating. This allows us to write the state 
vector $\ket{\psi}$ as the direct product of the even sector $\ket{\psi^e}$ and the 
odd sector $\ket{\psi^o}$:
\eq{
\label{evenodd}
\ket{\psi} =\ket{\psi^e}_e \otimes \ket{\psi^o}_o = \ket{\psi^e}_e 
\ket{\psi^o}_o }
provided the initial state can be decomposed accordingly. The dynamics in the
odd sector is essentially trivial, while many-body effects have to be considered
to capture the dynamics in the even sector, a topic that we address now.

\subsection{Many-body quantum dynamics with multi-mode coherent states}

The rationale behind the multi-mode coherent state (MCS) expansion
is as follows. The only source of non-linearity in Hamiltonian~(\ref{SB})
is the two-level system, and this non-linearity is transferred from
a single degree of freedom (the qubit) to a large number of degrees of
freedom (the modes of the waveguide). A first effect of this coupling
is to dress the two qubit states by displacing the oscillators,
as is clear from the $\sigma_z$ term in Eq.~(\ref{SB}). This picture,
which is only approximate when a single coherent state displacement is used,
becomes quantitavely exact for the many-body ground state when superposing
a small set of coherent states~\cite{bera_generalized_2014}.
Regarding the quantum dynamics, an input coherent state (as is relevant
in our description of the scattering problem) remains stable only when 
turning the coupling to zero (classical-like propagation).
At finite coupling, quantum fluctuations of the output field around the
dominant classical trajectory are again accounted for by the superposition 
of additional Gaussian states. The strategy is thus to write the state vector 
in the even sector as a coherent state expansion, also referred to 
in the following as the multi-mode coherent state (MCS) ansatz
\cite{bera_stabilizing_2014,bera_generalized_2014,snyman_robust_2015}:
\begin{equation}
\label{Psi}
|\Psi^e(t)\ra = \sum_{m=1}^{N_\mathrm{cs}} 
\Big[ p_m(t) |f_m(t) \ra | \ua \ra + q_m(t) |h_m(t) \ra | \da \ra 
\Big],
\end{equation}
where we have introduced the complex and time-dependent amplitudes $p_m(t)$ and
$q_m(t)$ for each qubit component, with $m$ an index that labels the states
used in the superposition. These multi-mode coherent states also occur
as two discrete sets of states (one for each qubit component):
\eq{
|f_m(t)\big> = \prod_{k=1}^{N_\mathrm{modes}} 
e^{[f_{k,m}(t) \aked -f_{k,m}^*(t) \ake]} |0\big> }
and similar for $|h_m(t)\big>$.
Due to the completeness of the coherent state basis on a discrete von-Neumann
lattice~\cite{boon_discrete_1978}, which naturally extends to the case of many
modes, this discrete decomposition can target in principle an arbitrary state of 
the full Hilbert space for $N_\mr{cs}\to\infty$. 
However, for a fixed choice of Gaussian states, this leads to
the unfathomable exponential cost that is typical of many-body quantum 
mechanics. The advantage of the MCS ansatz~(\ref{Psi}) lies in the variationally
optimized time-dependent displacements $f_{k,m}(t)$, which allows one to track with
high precision and low numerical cost the dynamics of the full state vector.

What is truly remarkable about such a multi-component multi-mode wavefunction is 
the relatively small number of coherent states $N_\mr{cs}$ that are necessary 
to capture both the static many-body ground 
state~\cite{bera_generalized_2014} and the complex dynamics resulting
from quantum quenches~\cite{gheeraert_spontaneous_2016}, even deep in the
ultra-strong coupling regime. The method works efficiently from the case of single
mode cavities~\cite{beraUnpublished,Cong,Clerk} up to the challenging situation
of an infinite continuum~\cite{SnymanKondo}.
As we will see later, addressing frequency conversion brings an additional difficulty 
in that non-linear emission signals are extremely faint when driving off resonance 
compared to the dominant elastic contributions, which requires very careful convergence 
of the numerics.

In principle, the exact Schr\"odinger dynamics, controlled by the Hamiltonian~(\ref{SB}), 
can be derived from the real Lagrangian density:
\begin{equation}
\mathcal{L}= \big<\Psi(t)|\frac{i}{2} \overrightarrow{\partial_t} 
-\frac{i}{2} \overleftarrow{\partial_t} 
- \mathcal{H}|\Psi(t)\big>,
\end{equation}
by applying the time-dependent variational principle~\cite{p._kramer_geometry_1981}, 
$\delta \int \mathrm{d}t \mathcal{L}=0$, upon arbitrary variations of the state 
vector~(\ref{Psi}) with respect to its set of variational parameters.
This minimization obviously provides Euler-Lagrange equations 
\eq{
\label{euler}
\frac{d}{dt} \frac{\partial \mathcal{L}}{\partial \dot{v}} =
\frac{\partial \mathcal{L}}{\partial v}}
for the set of variables 
$v=\{p_m,q_m,f_{k,m},h_{k,m}\}$, which can be solved by numerical
integration~\cite{burghardt_multiconfigurational_2003,yao_dynamics_2013,zhou_ground-state_2014,bera_dynamics_2016,gheeraert_spontaneous_2016}.
The detailed form of the dynamical equations is provided in
Appendix~\ref{AppDynamics}. A new numerical algorithm,
which supersedes the one proposed in Ref.~\cite{gheeraert_spontaneous_2016} and
allows to deal with up to thousands of modes, is presented in Appendix~\ref{algo}.

\subsection{General coherent state scattering formalism}
\label{formalism}

Now that we have obtained exact dynamical equations for the time evolution under 
the spin-boson Hamiltonian~(\ref{SBinit}), we need to prepare our initial state 
in order to perform scattering simulations according to the scheme in
Fig.~\ref{waveguide}. The generic difficulty is that the qubit is dressed
non-perturbatively by a cloud of photons~\cite{leggett_dynamics_1987,peropadre_2013,bera_generalized_2014,
sanchez_inelastic_2014,snyman_robust_2015,diaz_polaron_2016} in the ultra-strong 
coupling regime, so that this ground state assumes a many-body character.
Thanks to the MCS ansatz introduced in eq.~(\ref{Psi}), we can efficiently 
express the static ground state of the joint qubit and waveguide system in terms of multi-mode coherent 
states:
\begin{equation}
\label{Psi_GS}
|\Psi^\mrm{GS}\ra = \sum_{m=1}^{N^{\mrm{GS}}_{\textrm{cs}}} 
p^\mrm{GS}_m \Big[ |f^{\mrm{GS}}_m \ra_e | \ua \ra - |-f^{\mrm{GS}}_m \ra_e | \da \ra 
\Big]|0\ra_o,
\end{equation} 
where we enforced the $\mathbb{Z}_2$ spin-symmetry of the spin-boson Hamiltonian (\ref{SB}) to simplify
the expression. We have also used the fact that the odd modes do not interact with 
the qubit, so that the ground state displacements include only even modes in
Eq.~(\ref{Psi_GS}), and the odd modes are placed in the vacuum state.

By implementing numerically a variational optimization
\cite{bera_generalized_2014, bera_stabilizing_2014}, one can determine the set
of weights $p^\mrm{GS}_m$ and displacements $f^{\mrm{GS}}_{k,m}$, and thus obtain a 
nearly exact result for the ground state up to negligible numerical error. Conveniently, 
only a small number of coherent states $N^{\mrm{GS}}_{\textrm{cs}}$, typically less 
than 10, are required in the realistic domain of parameters of the spin-boson model. 

The second step in the scattering picture of Fig.~\ref{waveguide} is to include
a wavepacket (with arrow pointing inward) impinging on the dressed ground state. 
We will work in what follows with a single coherent state pulse as input, which 
is realistic in terms of the classical sources used in actual experiments.
Let us denote $z_k$ the displacement of the incoming wave-packet in mode $k$
of the physical waveguide, and $z_x$ its Fourier transform to real space:
\begin{equation}
\label{FT}
z_{x}=\int_{-\infty}^{\infty}\frac{dk}{\sqrt{2\pi}}e^{+ikx}z_{k}.
%\quad\textrm{and}\quad
%z_{k}=\int_{-\infty}^{\infty}\frac{dx}{\sqrt{2\pi}}e^{-ikx}z_{x}.
\end{equation}
We choose to use here a Gaussian-shaped wavepacket 
\begin{equation}
\label{zinput}
z_{k}=\sqrt{\bar n} \left(\frac{1}{2\pi\sigma^{2}}\right)^{\frac{1}{4}}
e^{-\frac{(k-k_{0})^{2}}{4\sigma^{2}}}e^{-i(k-k_{0})x_{0}}e^{-ik_{0}x_{0}/2},
\end{equation}
corresponding to a signal initially centered around position $x_{0}$ in the
waveguide, with mean wavenumber $k_{0}$, spatial extent $1/\sigma$, and 
total intensity corresponding to $\bar{n}$ photons on average, as illustrated 
on Fig. \ref{waveguide}.
The associated real space wavepacket is then
\begin{equation}
z_{x}=\sqrt{\bar n} \left(\frac{2\sigma^{2}}{\pi}\right)^{\frac{1}{4}}
e^{-(x-x_{0})^{2}\sigma^{2}}e^{+ik_{0}(x-x_{0})}e^{+ik_{0}x_{0}/2}.
\end{equation}
Note that these amplitudes are both normalized so that
$\int_{-\infty}^{\infty}dx|z_{x}|^{2}=\int_{-\infty}^{\infty}dk|z_{k}|^{2}
=\bar n$.

The even and odd parts of the incoming wavepacket are then defined strictly for 
$k>0$ as 
\begin{equation}
z_{k}^{e}=\frac{1}{\sqrt{2}}\left(z_{k}+z_{-k}\right)\quad\textrm{and}\quad
z_{k}^{o}=\frac{1}{\sqrt{2}}\left(z_{k}-z_{-k}\right).
\end{equation}
%which can be inverted to yield
%\begin{equation}
%z_{k}=\frac{1}{\sqrt{2}}\left(z_{k}^{e}+z_{k}^{o}\right)\quad\textrm{and}\quad
%z_{-k}=\frac{1}{\sqrt{2}}\left(z_{k}^{e}-z_{k}^{o}\right).
%\label{wavepacketbasis}
%\end{equation}
%Note that the first expression in Eq.~(\ref{wavepacketbasis}) gives the 
%right-going wave ($k>0$), while the second one is the left-going
%wave ($-k<0$).
%Given $z_{k}^{e}$ and $z_{k}^{o}$, the full wavepacket in real space $z_{x}$ can
%be reconstructed by
%\begin{equation}
%z_{x}=\int_{0}^{\infty}\frac{dk}{\sqrt{2\pi}}e^{+ikx}
%\frac{z_{k}^{e}+z_{k}^{o}}{\sqrt{2}}
%+\int_{0}^{\infty}\frac{dk}{\sqrt{2\pi}}e^{-ikx}
%\frac{z_{k}^{e}-z_{k}^{o}}{\sqrt{2}},
%\end{equation}
%where $x\in\mathbb{R}$ is the coordinate in the full waveguide.
Since even and odd modes commute, we can then define a displacement operator $D(z)$ for 
the initial incoming wave-packet which verifies:
\eq{
\label{WaveIn}
\ket{z^{e},z^{o}} &= D(z)\ket{0} = D(z^{e}) D(z^{o})\ket{0} \\
&= e^{\sum_{k>0}\left(z_{k}^{e}a_{k}^{e\dagger}-z_{k}^{e*}a_{k}^{e}\right)} 
e^{\sum_{k>0}\left(z_{k}^{o}a_{k}^{o\dagger}-z_{k}^{o*}a_{k}^{o}\right)} \ket{0}
\notag.
}
The final step in the initialization of the state vector is to combine 
the incoming wavepacket coherent state $z_k$ with the displacements
entering the full many-body ground state~(\ref{Psi_GS}). 
Straightforward calculations, shown in Appendix~\ref{AppState}, lead 
to the following explicit expression for the input state:
\begin{widetext}
\begin{eqnarray}
\label{PsiInit}
\ket{\Psi^\mr{IN}}
\nonumber
&=&\sum_{m=1}^{N^{\mrm{GS}}_{\textrm{cs}}} p^\mrm{GS}_m \Big[ \ket{\ua} 
e^{\frac{1}{2}\sum_{k>0}\left(z_{k}^{e}f_{k,m}^{\mr{GS}*}-z_{k}^{e*}f_{k,m}^\mr{GS}\right)}
 e^{\sum_{k>0}\left[\left(f_{k,m}^\mr{GS}+z_{k}^{e}\right)
a_{k}^{e\dagger}-\left(f_{k,m}^\mr{GS}+z_{k}^{e}\right)^{*}a_{k}^{e}\right]}\\
&& - \ket{\da} 
e^{\frac{1}{2}\sum_{k>0}\left(-z_{k}^{e}f_{k,m}^{\mr{GS}*}+z_{k}^{e*}f_{k,m}^\mr{GS}\right)}
e^{\sum_{k>0}\left[\left(-f_{k,m}^\mr{GS}+z_{k}^{e}\right) a_{k}^{e\dagger}
-\left(-f_{k,m}^\mr{GS}+z_{k}^{e}\right)^{*}a_{k}^{e}\right]}\Big]\ket{0}_e\ket{ z^o}_o.
\end{eqnarray}
\end{widetext}
The many-body scattering theory thus amounts to use state~(\ref{PsiInit})
as the initial condition for the dynamical equations of 
motion~(\ref{euler}) performed in the even sector, see Eqs.~(\ref{pevolution})-(\ref{fevolution}) 
for their full explicit form.
During the dynamics, as the incoming wavepacket impinges on the qubit, the necessary 
number of coherent states $N_\mr{cs}$ will sensibly grow from the initial value $N_\mr{cs}^\mr{GS}$
due to non-classical emission, therefore requiring to add extra coherent states 
to the state vector when needed (the procedure is detailed in Appendix~\ref{AppProtocol}). 
In the odd sector, which is completely decoupled from the qubit, the related
displacements are trivially evolving in time according to $i \dot{z}_k^o = 
\w_k z_k^o$, and a single coherent state is enough for the whole time-evolution.

After a given time $T$ long enough to ensure interaction of the wavepacket with the
qubit and subsequent decoupling of the two outgoing wavepackets from the
many-body cloud surrounding the qubit (in the reflection and transmission channel 
of the full 1D waveguide), one expects on general grounds (since the spin-boson
model is non-integrable with a realistic dispersion) a factorization of the final 
wavefunction as:
\begin{equation}
 \ket{\Psi(T)} = \ket{\Psi^{\mrm{GS}}} \otimes \ket{\Psi^{\mrm{OUT}}},
\label{PsiFinalTime}
\end{equation}
where $\ket{\Psi^{\mrm{GS}}}$ is the many-body ground state of the
spin-boson model and $\ket{\Psi^{\mrm{OUT}}}$ a many-body outgoing wavepacket
that contains a non-trivial decomposition of the emitted signal in terms of a large number 
of multi-mode coherent states (typically $N_{cs}^{\rm OUT} \sim 20-30$):
\begin{equation}
\label{PsiOUTk}
\ket{\Psi^{\mrm{OUT}}} = 
\sum_{m=1}^{{N^{\mrm{OUT}}_{\textrm{cs}} } } 
p^\mrm{OUT}_m
\prod_{k=1}^{N_\mathrm{modes}}
e^{[f_{k,m}^{\mr{OUT}} \aked -f_{k,m}^{\mr{OUT*}} \ake]} |0\big>.
\end{equation}
The extraction procedure for the outgoing weights $p^\mrm{OUT}_m$
and displacements $f_{k,m}^{\mr{OUT}}$ is given in Appendix~\ref{AppState}.
The factorization property (\ref{PsiFinalTime}) occurs because 
the spin-boson model (with a macroscopic number of modes) is a truly dissipative 
system, always showing a path for relaxation. In practice, this hypothesis
can be checked from the numerical calculations by observing that the dressed
qubit does not show correlations with the outgoing photons. Indeed, any observable 
of the qubit relaxes back to its initial equilibrium value at the end of the scattering 
protocol. Also, the nature of the scattered photons does not depend on how one
traces out the qubit density matrix. We now proceed to the analysis of the transmission 
and spectral properties of this scattered many-body wave-packet.

\section{Multiphoton inelastic scattering}
\label{Sec3}

\subsection{Elastic emission and high power saturation}

As a first illustration for our dynamical many-body scattering method, we investigate 
the elastic reflection as a function of the frequency and power of the incoming signal. This problem
is particularly challenging because of the combination of non-perturbative ultra-strong coupling 
with non-equilibrium effects that arise at finite input power. 
Ultra-strong coupling scattering at non-vanishing power has been addressed previously
with approximate techniques~\cite{lehur_kondo_2012,bera_dynamics_2016,shi_scattering,NIBA_2017} 
and with more advanced numerical methods~\cite{peropadre_2013,sanchez_inelastic_2014}. However,
systematic extraction of many-body scattering matrices has not been performed to 
our knowledge.

Our calculation scheme proceeds similarly to an experimental setup: the incoming Gaussian 
coherent-state wavepacket, shown schematically as the incoming distribution of photons in 
real space in Fig.~\ref{waveguide}, is initialized to the left of the qubit. 
The qubit is placed at position $x=0$ as seen from its sharply decreasing photonic 
cloud~\cite{snyman_robust_2015} which is present in both the input and output
ports, but remains statically bound to the central impurity.
After propagation towards the qubit and subsequent interaction, 
the photon flux decouples at long times, and is separated into a reflected
left going ($k<0$) signal and a transmitted right going ($k>0$) signal, both
shown with arrows pointing outward in Fig.~\ref{waveguide}. Note that in all the simulations made in this paper, 
we have considered the linewidth $\sigma$ of the wavepacket in $k$-space to be smaller 
than the qubit linewidth $\Gamma$ (in order to achieve high spectroscopic
resolution), but large enough to keep the simulations on a reasonable system size (typically 
we consider from $N_\mr{modes}=1000$ to $N_\mr{modes}=3000$ modes for the chain used 
in the even sector).
All calculations are done in the units of the plasma frequency $\omega_c$
as defined in the spectral density~(\ref{SpectralDensity}), and the wavepacket
linewidth appearing in Eq.~(\ref{zinput}) is taken as $\sigma=0.005\omega_c$, 
unless indicated otherwise.

We define the reflection and transmission coefficients in the following way:
\begin{equation}
\label{coefficients}
T=\frac{\sum_{k>0} \braket{\akd \ak}_{\mrm{out}}}{\sum_{k>0} \braket{\akd \ak}_{\mrm{in}}}\quad\textrm{and}\quad
R=\frac{\sum_{k<0} \braket{\akd \ak}_{\mrm{out}}}{\sum_{k>0} \braket{\akd \ak}_{\mrm{in}}},
\end{equation}
where we have denoted $\braket{\ldots}_{\mrm{in}}$ the average over the state vector
corresponding to the coherent incoming wave-packet before scattering, and 
$\braket{\ldots}_{\mrm{out}}$ the average over the many-body outgoing wave-packet after 
scattering. Both are obtained from the full state vector~(\ref{PsiOUTk})
by simply filtering out in real-space the polarization cloud associated with the ground 
state, as explained in Appendix~\ref{AppState}.
\begin{figure}[ht]
\includegraphics[width=1.0\linewidth]{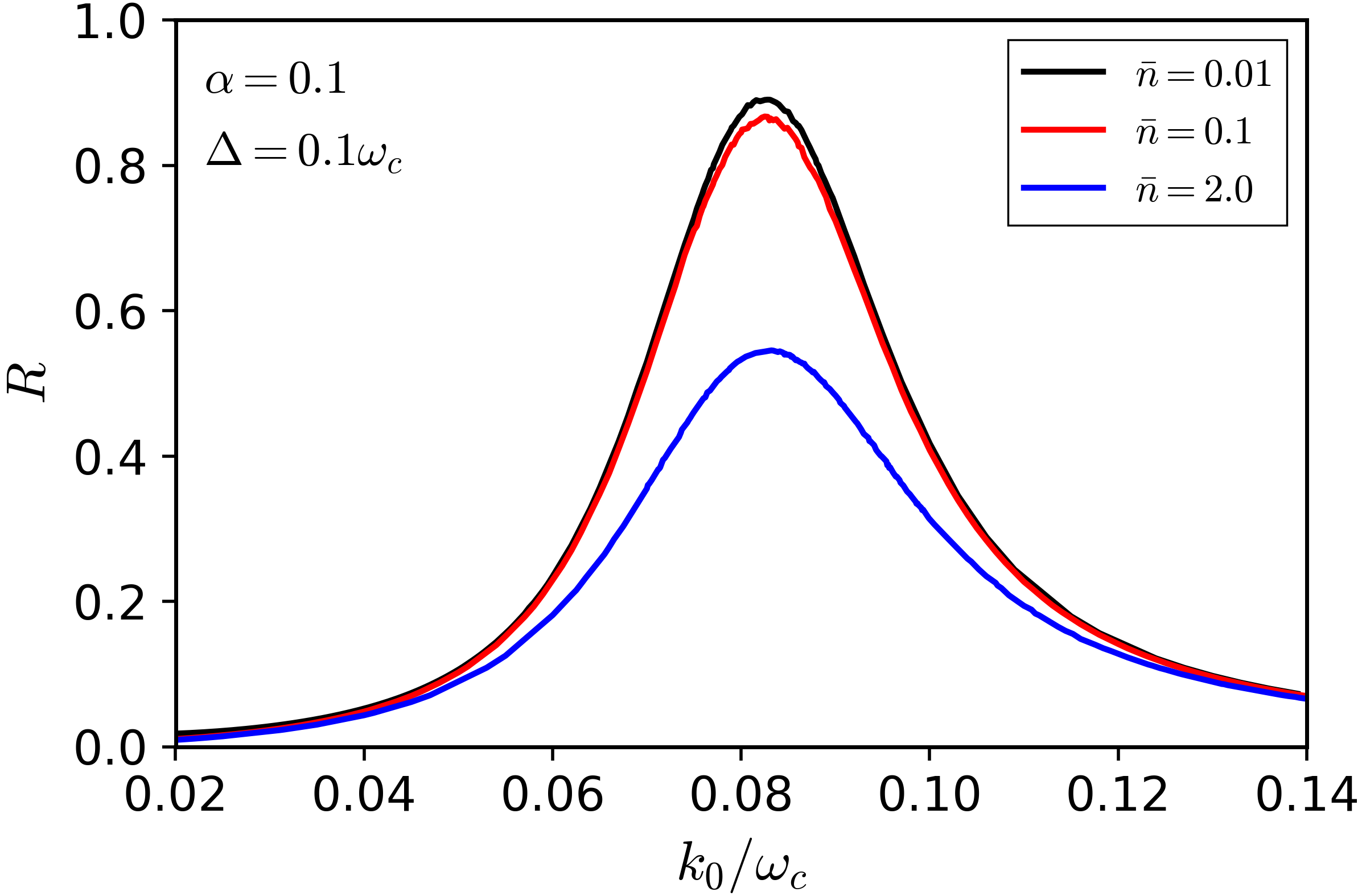}
\caption{
Saturation effects in the reflection coefficient 
Eq.~(\ref{coefficients}) as a function of incoming wavevector $k_0$, for three
different amplitudes $\bar n=0.01, 0.1, 2.0$ of the input, 
with a wavepacket width $\sigma=0.005\omega_c$.
These curves correspond to converged numerical data with up to $N_\mr{cs}=16$ coherent 
states in the MCS wavefunction~(\ref{Psi}). The bare qubit frequency is $\Delta=0.1\omega_c$ 
and the dimensionless light-matter interaction is $\alpha=0.1$, leading to a sizeable 
renormalized qubit frequency $\Delta_R\simeq0.08\omega_c$.}
\label{saturation}
\end{figure}

Results for different values of the incoming power are shown in Fig.~\ref{saturation}. 
The probability of reflection generally increases on resonance; indeed, for elastically 
scattered photons, interference effects cause almost complete reflection when exactly on resonance. 
For small values of the incoming power
($\bar{n}=0.01$ and $\bar{n}=0.1$), for which the initial coherent state wavepacket 
has a very small probability of containing Fock states with more than 1 photon,
one can note that the reflection only reaches $R \simeq 0.9$ at peak value. This
incomplete reflection of the photons arises from the finite linewidth of the
incoming wavepacket, and not from inelastic losses. Since our incoming 
Gaussian pulse is not perfectly monochromatic, the modes at the edge of a resonant incoming 
beam (centered at $k_0=\Delta_R$) are slightly off-resonant and do not get fully reflected 
by the qubit. 
Even in the present case of a relatively small light matter coupling $\alpha=0.1$, 
many-body effects due to the ultra-strong coupling are apparent 
in the reflection curve of Fig.~\ref{saturation}. First, a non-Lorentzian
asymmetric lineshape is obtained, with a high energy tail more prominent than at low energy.
In addition, we clearly observe a substantial renormalization of the qubit frequency 
$\Delta_R\simeq0.08\omega_c$ from its bare value $\Delta=0.1 \omega_c$.

For higher incoming power, one physically expects saturation effects to take
place, and these are clearly evidenced by the curve with average number
of photons $\bar{n}=2.0$ in Fig.~\ref{saturation}. We stress that 
converging such computations in the high power regime is quite challenging,
and approximate techniques such as a single coherent state truncation 
lead to uncontrollable noise levels, as found in previous
work~\cite{bera_dynamics_2016}. We show in detail in Appendix~\ref{AppConvergence}
that the reflection curve converges smoothly at $\bar{n}=2.0$ for about
$N_\mr{cs}=16$ coherent states in the MCS state vector~(\ref{Psi}). This is
also confirmed by a systematic control of the error, as done previously for
quantum quench protocols~\cite{gheeraert_spontaneous_2016}.

\subsection{Off-resonant frequency-conversion spectra}

\begin{figure}[t]
\label{off_res_Fock}
\includegraphics[width=1.0\linewidth]{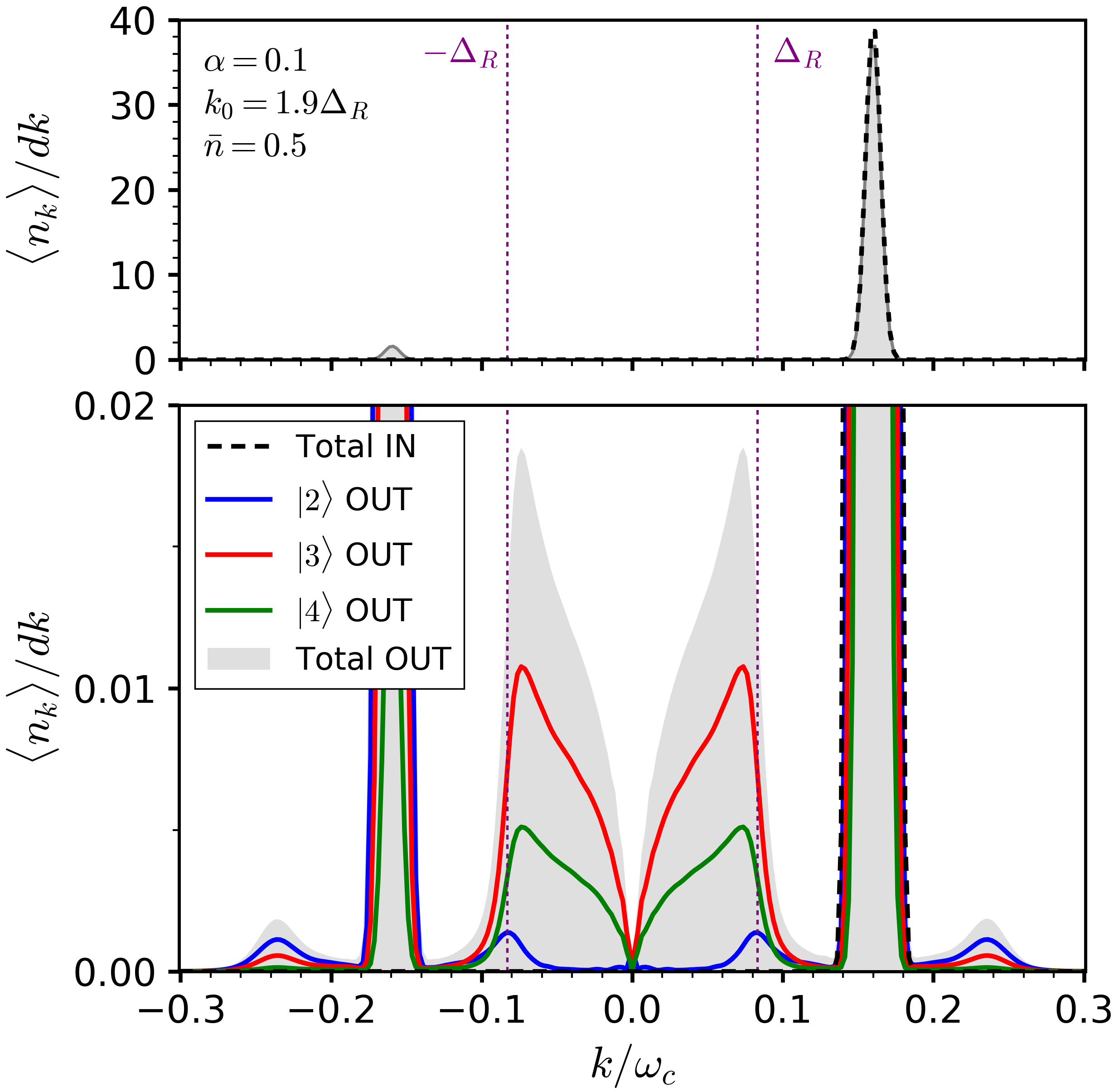}
\caption{
Mean density of photons at momentum $k$ in the
outgoing wave-packet (the incoming wavepacket is displayed as a dashed line); bottom panel is a zoom of top. 
The total outgoing signal (gray shaded)
is decomposed into Fock resolved excitations (in full lines) with $N=2$ (bottom
curve), $N=3$ (top curve) and $N=4$ (middle curve) photons, in order to highlight the processes of 
Fig.~\ref{diagrams}. The parameters for this simulation are the incoming wave-packet wave-number 
$k_0=0.16\w_c$ and linewidth $\sigma=0.005\w_c$, the mean photon number $\bar{n}=0.5$, the qubit bare 
energy $\Delta=0.1\omega_c$, and the coupling strength $\alpha=0.1$. Simulations were performed by 
considering a wavefunction containing $N_\mr{cs}=30$ coherent states, and $N_\mr{modes}=1200$ modes. 
}
\end{figure}

We now turn to analyzing the emitted radiation in 
the off-resonant case, in which the system is excited at a frequency $k_0$ above 
the renormalized qubit transition frequency $\Delta_R$. A typical inelastic spectrum is shown 
in Fig.~\ref{off_res_Fock}, here for $k_0 = 0.16\omega_c$, $\Delta_R=0.08\omega_c$, and an injected $\bar{n}$ of $0.5$. 
The stronger transmission relative to reflection (upper panel) simply reflects the 
off-resonant situation $k_0 \approx 2\Delta_R$, in agreement with the reflection curve in
Fig.~\ref{saturation}. The vertical scale is expanded in the lower panel, so that 
the inelastic contributions are made apparent at the foot of the large 
reflection and transmission elastic peaks located at $\pm k_0$. Note that the actual
linewidth of this elastic peak, set by $\sigma=0.005\omega_c$, is in fact much smaller 
than what the lower panel seems to indicate, because the maximum peak amplitude is 2000
times higher that the scale of the graph. 
The gray-shaded curve displays the expectation value of the total
number of outgoing photons $\braket {\akd \ak }_{\rm{out}}$ while the dashed
line indicates the total number of incoming photons $\braket {\akd \ak
}_{\rm{in}}$ centered around $k_0$. The first striking result is the broad spectrum of 
emission extending from the qubit frequency $\Delta_R$ all the way down to $k=0$.

The full lines display how the total outgoing photon contribution is
distributed among different Fock states $\ket{N}$ with photon number
$N=2,3,4$, allowing us to assess the nature of particle production. 
Our method to obtain these
photon-number resolved amplitudes by considering the probability 
of all the possible single- and multi-photon states for a given momentum $k$ is explained
in Appendix \ref{AppNumresolve}. Note that the majority of the inelastic
emission involves 3 and 4 photon contributions. Since the incoming average
photon number is only $0.5$, clearly substantial particle production is
occurring! Both the broad inelastic spectrum and particle production are
quintessentially ultra-strong coupling phenomenona.

\begin{figure}[t]
\includegraphics[width=1.0\linewidth]{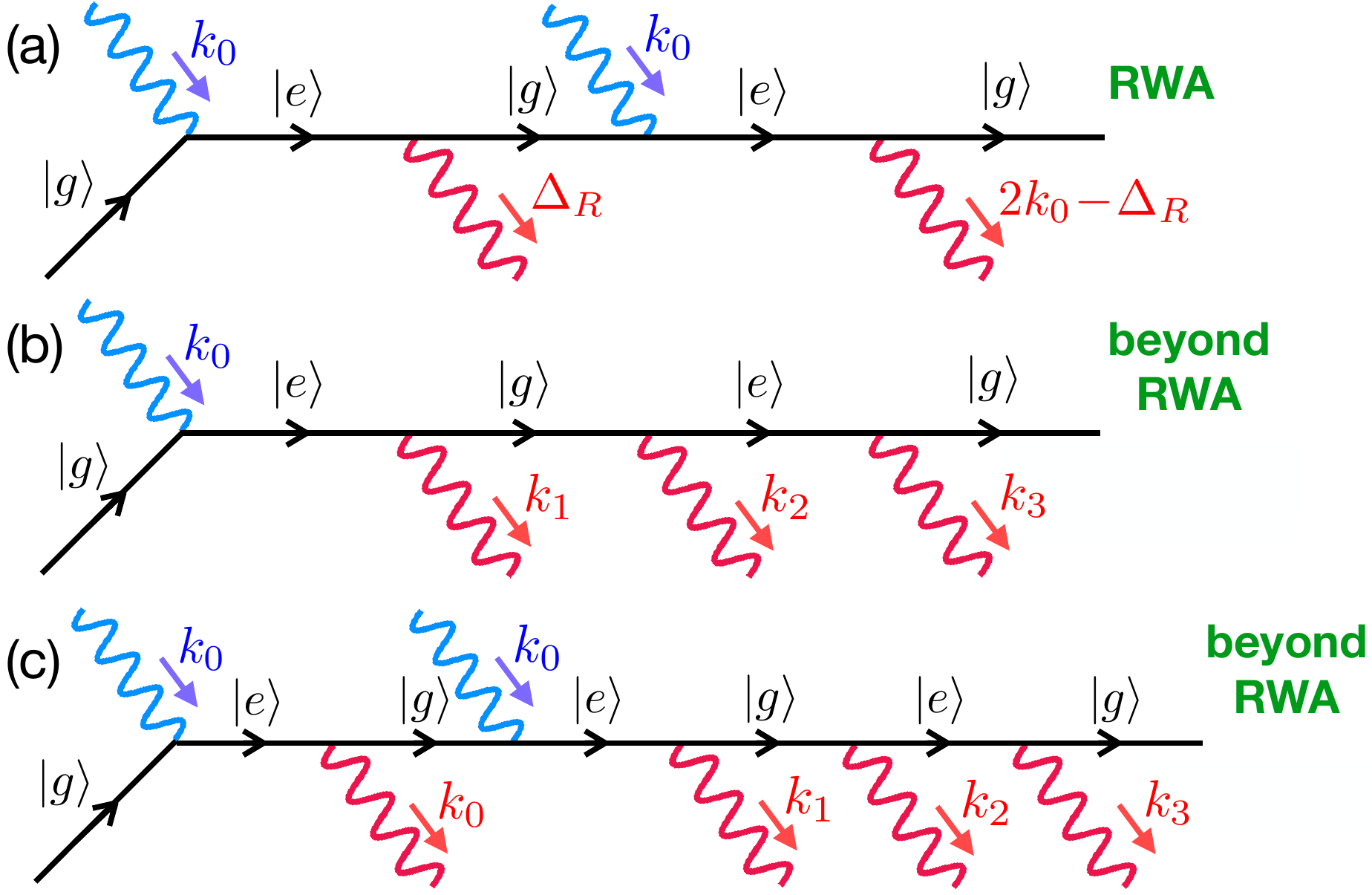}
\label{diagrams}
\caption{Diagrammatic representation of some non-linear photon processes occurring 
during scattering onto a two-level system. Panel (a) is a RWA frequency exchange process 
restricted at the two photon level, which shows two sharp emission lines at $\Delta_R$ and
$2k_0-\Delta_R$ (due to energy conservation).
Panel (b) shows the non-RWA one photon to three photon 
conversion~\cite{goldstein_inelastic_2013}, which leads to a broad emission
continuum, sharply peaked at the resonance $\Delta_R$ in the case of an
off-resonant drive. 
Panel (c) is a similar particle producing process, now with two input photons, 
one being down converted to three photons, the other photon being elastically scattered.}
\end{figure}

For an initial understanding of the various contributions to this spectrum, we consider
a schematic diagrammatic perturbation theory as shown in Fig.~\ref{diagrams}. 
For two incoming photons, an inelastic RWA process can occur by distributing the 
total incoming energy $2k_0$ into a resonant photon at $\Delta_R$ and another at
$2k_0-\Delta_R$ as shown in panel (a). 
However, for a single incoming photon with momentum $k_0$, since the emission is still
maximum at the (renormalized) resonant qubit frequency $\Delta_R$,  an excess
energy of $k_0-\Delta_R$ must be distributed between two extra outgoing photons
(in order to properly relax to the ground state). The accessible
non-resonant states thus lead to the non-RWA 
3-photon emission process shown in panel (b).
In general, the two extra photons that are produced are not 
resonant, and the amplitude of the total process is sizable only because of the
ultra-strong coupling regime. Indeed, the elastic reflection curve of
Fig.~\ref{saturation} is spectrally very broad, and emission does not
necessarily occur strictly on resonance.

The non-RWA nature of the particle production process is obvious from the non-conservation of
excitations: the middle outgoing arrow in Fig.~\ref{diagrams}(b) corresponds to 
the \textit{emission} of a photon upon
\textit{excitation} of the two-level system (instead of the usual de-excitation).
Four photon production is also displayed in panel (c) for an input state with two
photons. In this case, one input photon is elastically scattered at $k_0$, while
the second input photon splits into three photons similar to the process in
panel (b). Since the RWA $2\to2$ process in panel (a) and the non-RWA $2\to4$
process in panel (c) come at the same order in the input power, they can be used
to directly compare the relative strength of RWA and non-RWA processes. 
All three processes of Fig.~\ref{diagrams} are clearly observed in the spectrum
shown in  Fig.~\ref{off_res_Fock}, as the emission amplitude is decomposed
into photon number states $N=2,3,4$. In view of the wide use of the RWA in the
quantum optics context, the main surprise in these results (to be discussed in more 
detail below) is that non-RWA processes strongly dominate in amplitude the 
RWA processes.
\begin{figure}[t]
\label{off_res_alpha}
\includegraphics[width=1.0\linewidth]{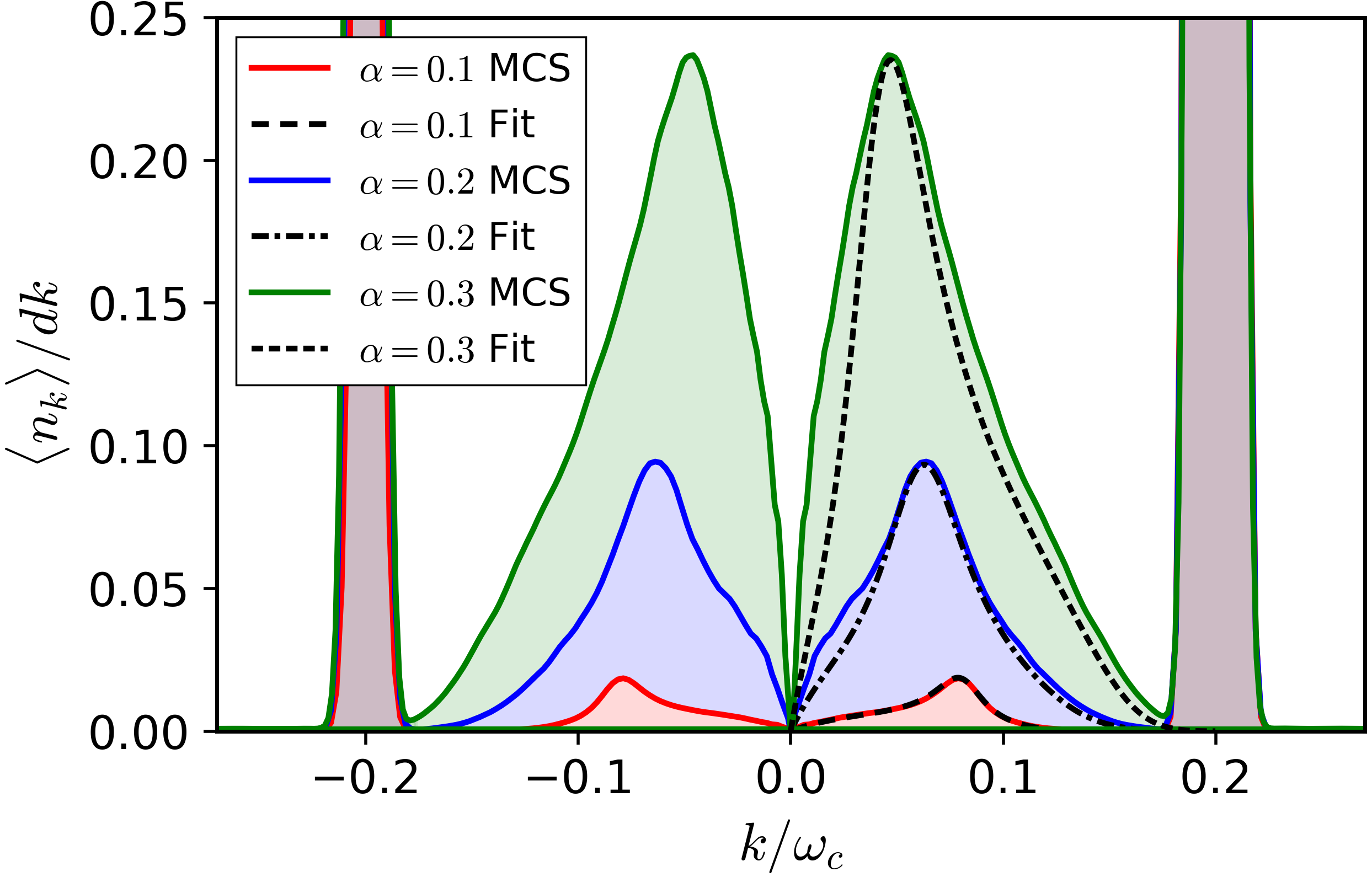}
\caption{Frequency conversion spectrum by going deeper in the ultra-strong 
coupling regime ($\alpha=0.1,0.2,0.3$, bottom to top curves) for an 
off-resonant incoming wavevector $k_0=0.2\w_c$.
Parameters are otherwise taken as in Fig.~\ref{off_res_Fock}. 
Although the resonance frequency $\Delta_R$ and inelastic linewidth $\gamma_R^\mr{inel}$
were fitted, one observes increasing deviations to the fitting 
formula~(\ref{CompGoldstein}) at larger $\alpha$. Enhanced scattering 
of low-energy modes, precursor of the Kondo regime, originate from non-perturbative 
many-body corrections beyond the lowest order perturbation theory of
Ref.~\cite{goldstein_inelastic_2013}.} 
\end{figure}

Some of the off-resonant processes were previously predicted perturbatively by Goldstein 
et al.~\cite{goldstein_inelastic_2013} in the $\alpha\to0$ limit and at the
Toulouse limit, and we are able to characterize quantitatively the non-linear emission 
for the first time at finite $\alpha$ values, as seen in Fig.~\ref{off_res_alpha}. 
The main effect brought by stronger coupling is a further renormalization
of the spontaneous emission line $\Delta_R$ down to lower values, as well
as a global increase of the probability for inelastic conversion.
Interestingly, we find that the perturbative formula~(\ref{CompGoldstein}) cannot 
quantitatively describe our data anymore in this regime, even when allowing to
fit the inelastic linewidth. Perturbation theory thus fails to capture the pile-up 
of low-energy photons found in the numerical simulations,
which signals the approach to the incoherent Kondo regime, in which the qubit 
resonance is fully washed out.
A detailed study of non-linear spectra as a function of incoming momentum is given in
Appendix~\ref{GoldsteinApp}.

\begin{figure*}[htb]
\label{3dplots}
\includegraphics[width=1.0\linewidth]{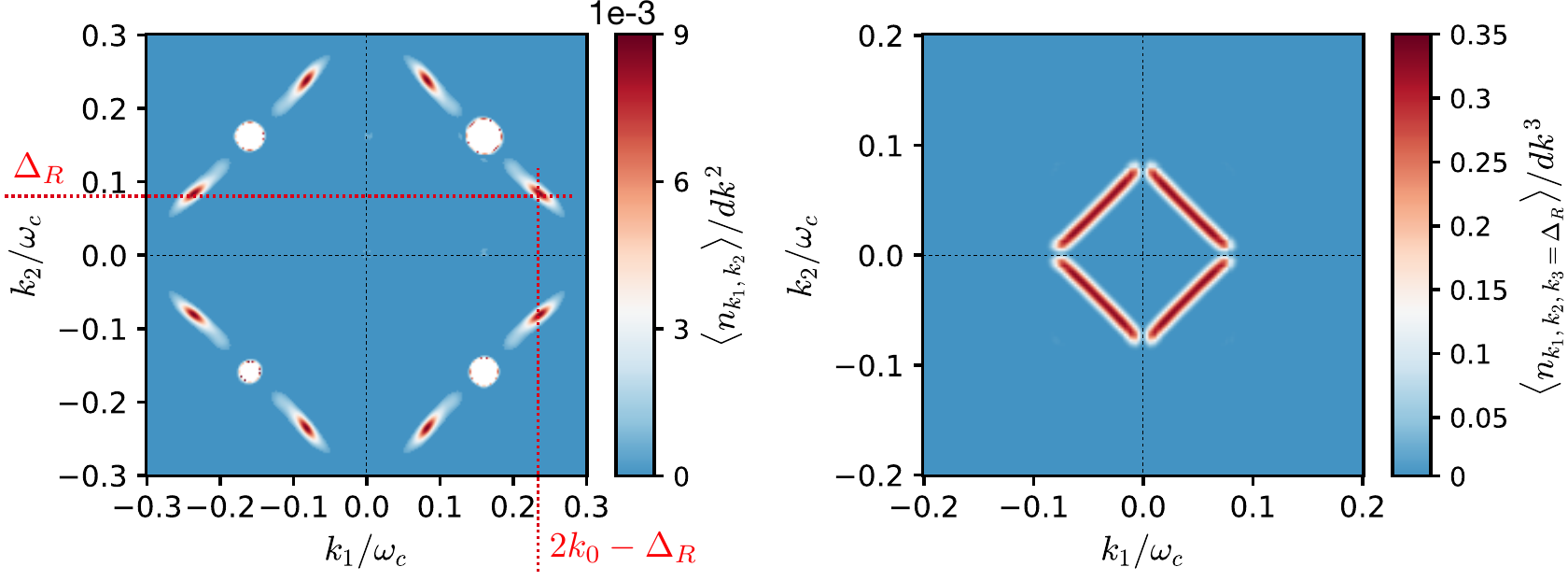}
\caption{Left panel: probability distribution of the two-photon states
$|\alpha_{k_1,k_2}|^2/dk^2$ corresponding to the 2-photon curve in Fig.
\ref{off_res_Fock}. The clear inelastic sidebands correspond to
frequency exchange between 2 photons.
 Right panel: probability density distribution
$|\alpha_{k_1,k_2,k_3=\Delta_R}|^2/dk^3$ of the three-photon states in which one
photon is at the resonance frequency 
$k_3 \approx \Delta_R$, corresponding to the 3-photon
curve in Fig. \ref{off_res_Fock}. The emission continuum associated to 
particle production is revealed by the diamond-shaped line. 
}
\end{figure*}
\subsection{Particle production processes}
We now investigate more precisely the photonic content of the emitted radiation 
in the inelastic channel. Let us start with the 2-photon
particle-conserving RWA contribution (bottom full line $N=2$ in Fig.~\ref{off_res_Fock}) forming two 
lobes symmetrically arranged around the main elastic peak (at $k_0=0.16\w_c$).
The lowest energy lobe is centered around $k \simeq 0.08\w_c \simeq 
\Delta_R $ corresponding to the spontaneous reemission of the qubit, while the
high energy lobe is located around $k \simeq 0.24\w_c\simeq 2k_0-\Delta_R$, as
expected from energy conservation [panel (a) in Fig.~\ref{diagrams}].
A closer view into this two-photon joint emission process is given
by the complete two-photon probability distribution $|\alpha_{k_1\, k_2}|^2$
that is plotted in the top panel of Fig.~\ref{3dplots} (see Appendix~\ref{AppNumresolve} for
details).
The main 2-photon elastic peaks are the white disks located at $[\pm k_0,\pm k_0]$, 
that have been cut off in order to magnify the small inelastic contributions. 
From the lateral inelastic peaks, one can immediately read-off the two-photon 
frequency conversion process in which two photons with energy $k_0$ redistribute their energy
into one photon with momentum $\Delta_R$ and another with energy $2 k_0 - \Delta_R$.

The inelastic spectrum originating from the conversion of a single incoming 
photon into three outgoing photons, with probability $\braket{n_k}_\mr{3photon}$
of measuring one of these photons at energy $k$, is represented by the middle 
full line in Fig.~\ref{off_res_Fock}. This inelastic lineshape presents quite
unusual features: a sharp resonance at the qubit frequency $\Delta_R$, a broad 
continuum extending from zero energy up to the foot of the elastic peak, and a
small lobe at the same energy $2k_0-\Delta_R$ as the previous two-photon 
conversion process. The latter is easily understood as an input of three photons
with momentum $k_0$, out of which one photon is elastically scattered, while the
other two are RWA frequency converted to $\Delta_R$ and $2k_0-\Delta_R$ (similar
to the previous $2\to2$ RWA process). We have checked that this $3\to3$ RWA
process becomes relatively weaker in amplitude as the input power $\bar{n}$ is
turned down, and is indeed associated to a three-photon input.

The broad low-energy continuum is readily explained by the 1-photon to 3-photon
non-RWA conversion process shown in panel (b) of Fig.~\ref{diagrams}.
This interpretation is backed up by studying in the right panel of Fig.~\ref{3dplots} 
the probability distribution $|\alpha_{k_1 k_2 k_3=\Delta_R}|^2$ of 3-photon outgoing 
states for which one of the three outgoing
modes is resonant, $k_3=\Delta_R$. To understand this diamond-shaped pattern, one can 
observe that the process leading to the diagonal line in the top-left quadrant can
easily be parametrized as:
\eq{ \ket{k_0} \rightarrow \ket{ \Delta_R} \ket{\Omega}\ket{ k_0 - \Delta_R -
\Omega } \; \mrm{with} \; \Omega \in [0:k_0-\Delta_R] \notag .} 
which basically expresses the conservation of energy between the input and the
output. Note that in the on-resonant situation (or for a drive at frequency below 
$\Delta_R$), all emitted photons present energies below the qubit frequency.

The next section investigates how the complete inelastic emission spectra compare with
the standard RWA prediction in quantum optics. This comparison will provide not
only a benchmark of our simulations, but also several physical signatures that
cannot be captured without the inclusion of particle production processes.

\section{Success and failure of the RWA for non-linear emission}
\label{sec4}

\subsection{RWA inelastic conversion}

To highlight particle production that arises at ultra-strong coupling, we 
now compare our MCS simulations to a direct treatment within the RWA, an approximation 
which conserves the number of excitations. Transport under the RWA is obtained in 
the framework of input-output theory.
Within the RWA, it is convenient to work in the basis that diagonalizes the
qubit. After applying the rotating wave approximation to the Hamiltonian
\eqref{SBinit} and assuming a frequency-independent coupling constant 
$g_k= \sqrt{\alpha \Delta_{R}} $, one finds that the system is described by the
Hamiltonian
\begin{eqnarray}
H &=&\frac{1}{2} \Delta \sigma^{z}
+ \int d \omega \,\, \frac{g}{2} \big[ \sigma^{+} (r_{\omega} + l_{\omega}) +
h.c. \big]\\
\nonumber
&&+ \int d \omega \,\, \omega 
\big( r_{\omega}^{\dagger} r_{\omega} - l_{\omega}^{\dagger} l_{\omega} \big),
\end{eqnarray}
where $\sigma^{+}$ is the raising operator of the qubit and $r_{\omega}
(l_{\omega})$ is the annihilation operator for the right(left)-going mode of
frequency $\omega$. We adapt standard input-output theory for a monochromatic
input \cite{WallsMilburnQO08, FanPRA10, KocabasPRA12, LalumierePRA13} to our
case of an incoming wavepacket with finite energy resolution.
The input-output relation remains the usual one,
%\begin{equation}
%\label{IO}
$r_{\text{out}}(t) = r_{\text{in}}(t) - i \sqrt{\pi/2}\, g \, \sigma^{-} (t)$
%\end{equation}
and similarly for the left-going field $l_\text{in/out}$. This allows one to
find the properties of the outgoing field from a master equation for the qubit.
In this way the power spectrum is calculated through the first-order correlation
function $G^{(1)}(t_1,t_2) = \braket{a_{\text{out}}^{\dagger}(t_1)
a_{\text{out}}(t_2)}$ by a Fourier transform
\begin{equation}
S[\omega] = \frac{1}{2\pi} \int_{0}^{T} dt_{1} \int_{0}^{T} 
dt_{2} \, G^{(1)}(t_1,t_2) e^{i \omega (t_2-t_1)}.
\end{equation}

We assume that the qubit is located at $x=0$ while the input and output ends are
located at $x=-T/2$ and $T/2$ respectively ($c=1$). From the definition
\eqref{zinput}, we can write the wavepacket in frequency as 
\begin{equation}
z(\omega) = \sqrt{\bar n} \left(\frac{1}{2\pi\sigma^{2}}\right)^{\frac{1}{4}}
e^{-\frac{(\omega-k_{0})^{2}}{4\sigma^{2}}} e^{i (\omega-k_0) T/2},
\label{InputGaussian}
\end{equation}
through which the input coherent state is defined as $\ket{z^{+}} = \exp{\big[
\int dk\, z(\omega)\, r_{\text{in}}^{\dagger}(\omega) - h.c.\big]} \ket{0}$,
where $r_{\text{in}}^{\dagger}(\omega)$ is the standard monochromatic input
operator \cite{WallsMilburnQO08, FanPRA10, KocabasPRA12, LalumierePRA13} of
input-output theory.  The input operator describing our wavepacket then
satisfies
\begin{equation}
r_{\text{in}}(t) \ket{z^{+}} = \frac{1}{\sqrt{2 \pi}} A(t) e^{-i k_{0} t} \ket{z^{+}} 
\text{ and } l_{\text{in}}(t) \ket{z^{+}} =0 ,
\end{equation}
where $A(t) = \tilde{A} e^{-\sigma^2 (t-T/2)^2}$ with $\tilde{A} = \sqrt{2
\bar{n} \sigma} (2 \pi)^{1/4}$ is the change of driving amplitude on the qubit
with time as the Gaussian wavepacket passes by. 

A master equation for the qubit density matrix $\rho_{s}$ is then obtained by
transforming to the Schr\"odinger picture,
\begin{equation} \label{MasterE}
\begin{split}
\frac{\partial}{\partial t} \rho_{\text{s}} =& - i \big[ \frac{\delta}{2} \sigma^{z} 
+ g A(t) \sigma_{+} + h.c.\, , \rho_{\text{s}} \big] \\
&+ \pi g^{2} \big( \sigma_{-} \rho_{\text{s}} \sigma_{+} 
-\frac{1}{2} \{ \rho_{\text{s}}, \sigma_{+} \sigma_{-} \} \big),
\end{split}
\end{equation}
where a rotating frame given by $k_{0} \sigma^{z} /2$ has been used. Note that
decay rate is $\Gamma = \pi g^{2} = \pi \alpha \Delta_{\text{R}}$.  For the
reflected light, the power spectrum can be shown to be
\begin{equation}
S_{\text{L}}[\omega] = g^2 \int_{0}^{T} dt \int_{-t}^{T-t} d\tau 
\braket{ \sigma_{+}(t) \sigma_{-}(t+\tau) } e^{i (\omega-k_{0}) \tau}; 
\end{equation}
two additional interference terms appear in the power spectrum for the
transmitted light and are not given here. The desired correlation function
$\braket{ \sigma_{+}(t) \sigma_{-}(t+\tau) }$ can be calculated through the
master equation \eqref{MasterE} and the quantum regression theorem
\cite{Carmichael93}.

\begin{figure}[ht]
\label{spectrum_comp}
\includegraphics[width=1.0\linewidth]{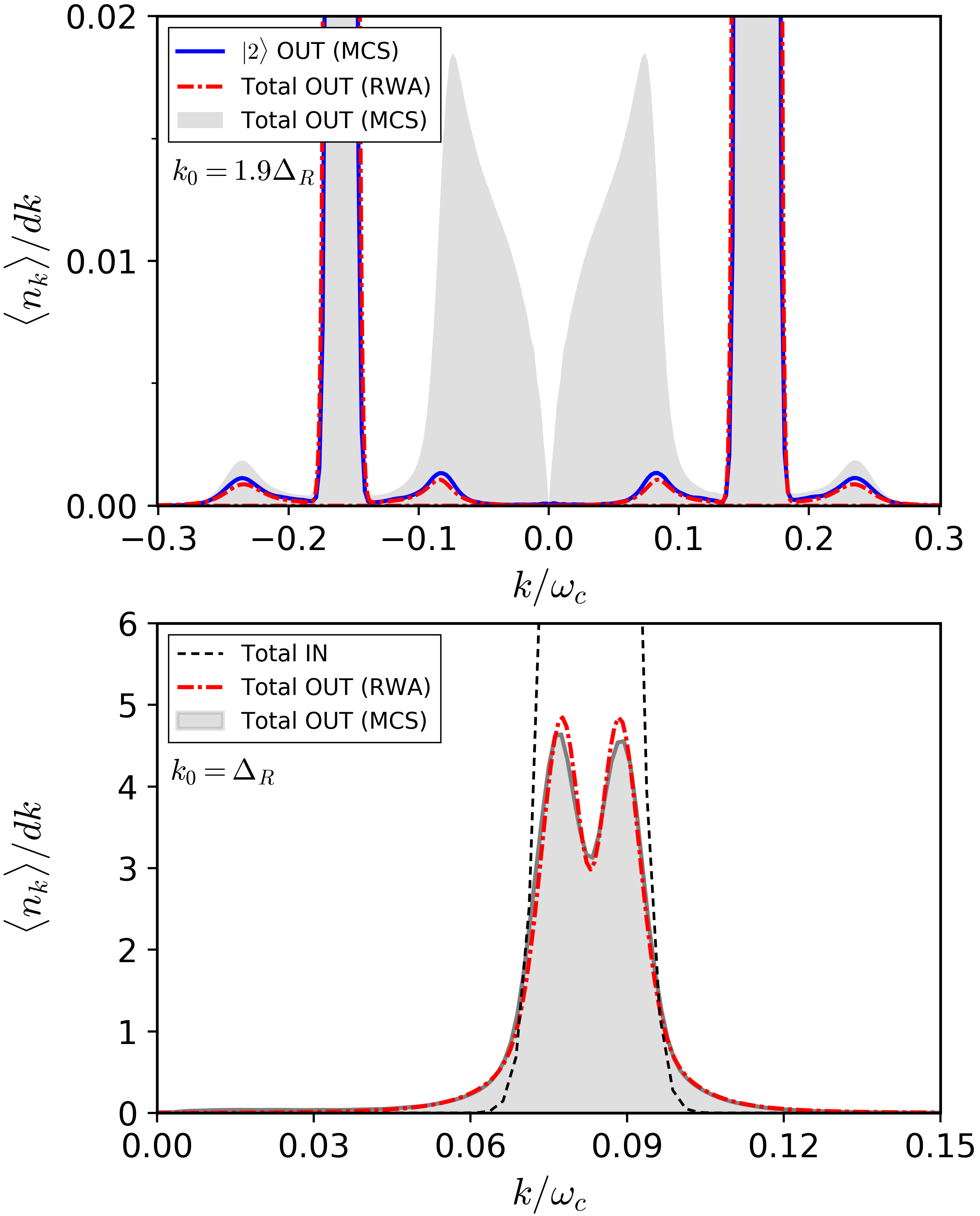}
\caption{
Comparison of MCS simulations to RWA input-output theory with regard to 
frequency conversion spectra for the off-resonant case (upper panel)
and on-resonant case (lower panel). The parameters are the same as in 
Figs.~\ref{off_res_Fock} and \ref{on_res}.
This confirms the previous
interpretation that RWA inelastic processes dominate only on-resonance, and miss
the main contributions to the off-resonant signal.} 
\end{figure}
Comparison to the RWA power spectrum in the off-resonant and resonant cases is
shown in the upper and lower panels of Fig.~\ref{spectrum_comp} respectively.
In making this comparison, we used as input to the RWA calculation the
numerically found renormalized level spacing and width, $\Delta_R$ and 
$\Gamma_R = \pi\alpha\Delta_R$, as this is essential to get the elastic peak 
correctly.
The dominant inelastic process within the RWA is the scattering of two incoming
photons into two outgoing photons, see panel (a) in Fig.~\ref{diagrams}. One sees that this
process explains most of the total scattered spectrum in the resonant case.  
Indeed, the lower panel in Fig.~\ref{spectrum_comp} shows that the RWA
and numerical MCS results are nearly identical on the scale shown. In particular
both the overall width and shape of the inelastic power spectrum agree well.
However, it is clear that the RWA prediction is only a small fraction of the total 
inelastic scattering in the off-resonant case (upper panel in
Fig.~\ref{spectrum_comp}), as particle production leads to qualitatively 
different and much larger cross-sections. Thus for these parameters, the RWA fails 
badly, even though the coupling constant $\alpha = 0.1$ is not very large.

\subsection{Temporal correlations associated to particle production}

It is interesting to study photon number temporal correlations, a standard measure of
non-linearities, but now in light of the large inelastic effects that we uncovered 
in the ultra-strong 
coupling regime. We have computed the photon-number autocorrelation function $g_2(\tau)$ 
of the reflected signal ($x<0$, $k<0$), defined by
\eq{
\label{g2}
g_2(\tau) = \frac{\braket{a^\dag_{x} a^\dag_{x+\tau} a_{x+\tau} a_{x} }}
{\braket{a^\dag_{x+\tau} a_{x+\tau}}\braket{a^\dag_{x} a_{x}}},
}
where $x$ is a point within the left-going wavepacket, such that both $x$ and
$x+\tau$ are within the wavepacket. In principle, $g_2(\tau)$ also
depends on $x$, but this dependence is weak provided the wavepacket is almost
monochromatic, and the location $x$ is taken deep within the outgoing photon
wavepacket. Details of the computation in the context of an MCS expansion are
given in Appendix \ref{App-g2calc}.
\begin{figure}[ht]
\label{anti}
\includegraphics[width=1.0\linewidth]{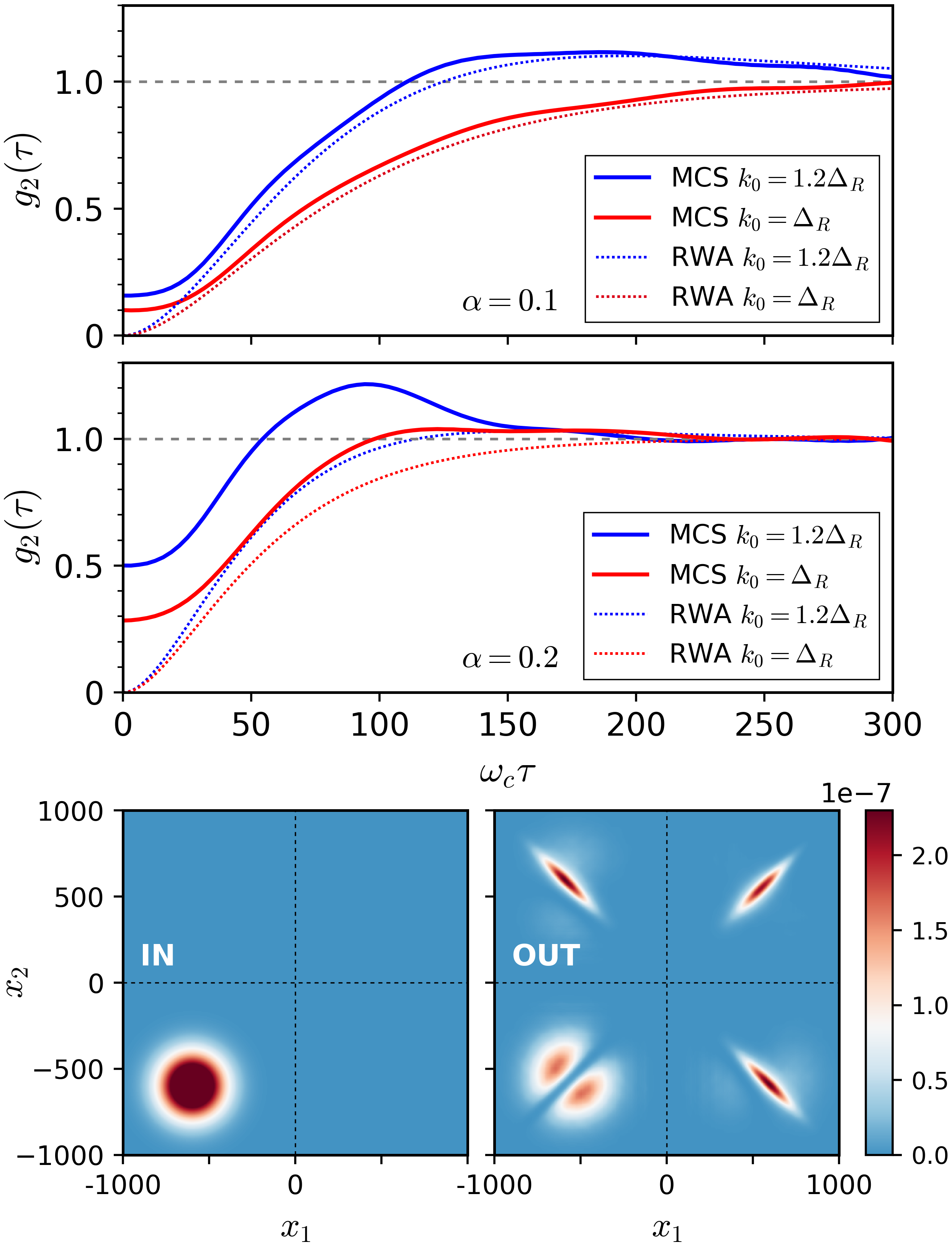}
\caption{Second order correlation function $g_2(\tau)$ in reflection,
at $\alpha=0.1$ (upper panel) and $\alpha=0.2$ (middle panel).
The dip at $\tau=0$ is a standard signature of anti-bunching, but multiple 
photon emission seen in Fig.~\ref{off_res_Fock} at ultra-strong coupling 
leads to an incomplete cancellation, $g_2(0)>0$. In the off-resonant case
($k_0=1.2\Delta_R$), particle production is enhanced 
relative to the single photon reflection, resulting in a stronger bunching 
($g_2(\tau)>1$) than predicted in the RWA.
The MCS simulations were performed with the same parameters as
in Fig.~\ref{on_res}, except for the stronger coupling $\alpha=0.2$,
$\sigma=0.004 \omega_c$ and a hard cut-off that was used (see Appendix \ref{App-g2calc}).
Bottom panel: Real space probability distribution of the two-photon states 
$|\alpha_{x_1,x_2}|^2/dx^2$ at the initial (left panel) and final (right panel) times 
of the simulation. One remarks the absence of reflection for two photons arriving at 
the same time on the qubit, as seen by the dip within the probability distribution 
located in the bottom left quadrant of the right panel.
}
\end{figure}

We find that temporal correlations are a very sensitive measure of ultra-strong
coupling effects. In the resonant case (see the top panel of Fig.~\ref{anti}), the
correlations are typical of single photon emission.
The comparison to the RWA is globally quantitative, as expected from the previous 
agreement in the inelastic spectrum on-resonance (small oscillations at long
time in $g_2(\tau)$ reflect the improper convergence of our MCS numerics near the edges 
of the outgoing wavepacket). In disagreement
with the RWA however, we notice that the numerical data shows partial
antibunching at zero delay, $g_2(0)>0$, signaling the production of particles,
as was revealed by the low energy spectrum in Fig.~\ref{on_res}.  Thus particle
production leads to physical effects that are potentially observable experimentally 
even when on resonance. This offset, which is zero in the RWA, is found to 
increase with $\alpha$ (see the upper middle of Fig.~\ref{anti} for $k_0=\Delta_R$).
The incomplete cancellation here can be readily interpreted as a probability of
emitting many-photon states due to frequency down-conversion. Even more striking
is the appearance of a large bunching signal at intermediate times in the
off-resonant case (see the middle panel of Fig.~\ref{anti} for $k_0=1.2\Delta_R$), 
which was not reported to our knowledge for the radiation 
of a \textit{single} level qubit (bunching can be observed in spontaneous emission 
from multilevel atoms~\cite{Sanchez16}, due to a simpler cascade effect~\cite{gasparinetti_2017}, 
or from multiqubit systems \cite{ZhengPRL13,Pletyukhov14,FangPRA15}). 
Here, bunching originates from the single-shot emission of three photons by the 
two-level system, a property that is only allowed at ultra-strong coupling.
The bunching signal becomes sizable in the off-resonant case, even though
the particle production is comparable to that in the resonant case, because the
reflection amplitude for single photon emission is reduced. 

As a nice illustration of the partitioning of the incoming beam by the two-level
system, we show in the bottom panels of Fig.~\ref{anti} the real space
probability distribution of the two-photon states $|\alpha_{x_1,x_2}|^2/dx^2$ at
the beginning and at the end of the time evolution. These results were obtained
in the on resonant case with $\alpha=0.1$, by Fourier transforming to real space
the $k$-space displacements. One can clearly see within the reflected signal 
(bottom left quadrant in the right panel) a deep trench on the
diagonal $x_1=x_2<0$ with vanishing photon content (the incoming coherent state
is shown in the bottom left panel for comparison). Two photons impinging
simultaneously on the qubit have thus very low likelihood 
of both being reflected. This provides a direct visualization 
of photon anti-bunching, which arises because a single emitter can only reflect one photon at a time. 
%Two photons impinging simultaneously on the qubit have thus very low likelihood 
%of both being reflected at once, leading to the conclusion that the reflected signal 
%is anti-bunched, as was evidenced by the $g_2$ function.

\section{Conclusion and perspectives}
\label{Sec5}

In this work we have developed a powerful methodology, namely the MCS technique, based on multi-mode
and multi-configuration coherent state wavefunctions, to address many-body 
scattering properties of a two-level system that is embedded in a waveguide in 
the regime of ultra-strong coupling. This problem is intrinsically non-perturbative 
in nature due to the large production of particles, and cannot be reliably addressed 
by standard methods in quantum optics.

Our main finding is that excitation-preserving processes, described by
the rotating wave approximation (RWA), 
dominate the inelastic spectrum only in the resonant
situation. In contrast, when the frequency of the incoming photons is larger than 
the renormalized transition frequency of the two-level system, particle production
becomes very favorable and dominates the inelastic signal. 

We have been able to characterize precisely the output field, by decomposing the reflected and
transmitted photon wavepackets into Fock states, and also by computing temporal
correlations. 
The main results are as follows. 
(i)~The process by which one photon is absorbed and three photons are emitted dominates 
in the off-resonant low power limit and leads to a broad spectrum of emission
extending from zero frequency to the renormalized qubit frequency.
(ii)~Even in the resonant case, while the dominant inelastic emission near the resonant
frequency is captured by the rotating wave approximation, there is still a broad
spectrum of weak inelastic transmission produced by the counter-rotating terms.
(iii)~The correlation function $g_2(\tau)$ in reflection is a sensitive measure
of ultra-strong coupling physics. In particular, particle production implies
that it needs not vanish at zero delay, $g_2(0)>0$, and that it shows a
strong bunching effect at a delay of order the inverse lifetime. 
(iv)~Finally, we have found that perturbative predictions for the inelastic 
response~\cite{goldstein_inelastic_2013} cannot be used simply by renormalizing 
the bare qubit resonance frequency and linewidth when the coupling becomes
ultrastrong. A more consistent theory including self-energy effects should be 
developed for the future.

All our quantitive predictions have relevance for the ongoing experimental effort in 
pushing waveguide quantum electrodynamics to the ultra-strong coupling regime. 
The connection to future experiments
opens in addition various research directions. One important issue is that
superconducting qubits are rarely operated as truly perfect two-level systems.
Reducing the non-linearity of the qubit is typically important to minimize the
effect of random noise from the circuit, but this strongly diminishes of course
the amplitude of the interesting non-linear signals. Thus, extending our
methodology to fully realistic superconducting quantum circuits will be crucial
to address whether particle-production can be sizeable in practice.

The ability of the multi-coherent state method to deal
naturally with coherent state pulses and open environments is also relevant for
the large interest in quantum manipulation within complex architectures. 
It would thus be very useful to adapt techniques from signal treatment in order to 
numerically optimize the quantum evolution of the displacements that are used to 
simulate the Schr\"odinger dynamics of the complete system.
Such developments will certainly be useful, because the description of strongly driven
open quantum systems is a very important topic currently. Based on the physical
artifacts that we can observe in our simulations of the scattering problem 
when the wavefunction is far from being converged, we suggest
that the description of non-linear effects in quantum circuits for arbitrary pulse 
sequences is a very delicate subject that has to be examined with advanced and reliable 
many-body techniques.

\section*{Acknowledgments}
We thank Manuel Houzet and Izak Snyman for stimulating discussions.
N. G. acknowledges support from the Fondation Nanosciences de Grenoble.
S. B. acknowledges support from DST, India, through Ramanujan Fellowship
Grant No. SB/S2/RJN-128/2016.
H. U. B. acknowledges travel support from the Fondation Nanosciences de Grenoble under 
RTRA contract CORTRANO. The work at Duke was supported by U.S. DOE, Division of Materials 
Sciences and Engineering, under Grant No. DE-SC0005237.
S. F. and N. R. are supported by the ANR contract CLOUD (project number
ANR-16-CE24-0005).

\appendix

\section{Technical aspects of the simulations}
\subsection{Dynamics of the MCS state vector}
\label{AppDynamics}
The multi-mode coherent state decomposition~(\ref{Psi}) leads to compact
Euler-Lagrange equations~(\ref{euler}) that determine the full 
quantum dynamics the spin-boson model~(\ref{SBinit}):
\begin{eqnarray}
\hspace{-0.7cm} P_j &=& \sum_m \left(\dot{p}_m-\frac{1}{2} p_m \kappa_{mj}\right) M_{jm}, 
\label{pevolution} \\
\hspace{-0.7cm} F^k_j  &= & \sum_m\left( p_m \dot{f}_{k,m}+\left(\dot{p}_m-\frac{1}{2} p_m
\kappa_{mj}\right)f_{k,m}\right)M_{jm},
\label{fevolution}\\
\hspace{-0.7cm} \kappa_{ij} &=& \sum_k \left(\dot{f}_{k,m}f^{\star}_{k,m}
+\dot{f}^{\star}_{k,m}f_{k,m}-2f^{\star}_{k,j}\dot{f}_{k,m}\right).
\end{eqnarray}
Here, 
$ M_{ij}= \braket{f_i|f_j}=
{\rm e}^{ -\frac{1}{2} \sum_k  [|f_{k,i}|^2 + |f_{k,j}|^2 - 2 f_{k,i}^*
f_{k,j}] }$ corresponds to the overlap between two multi-mode coherent states, and
arises in the equations because of the over-completeness of the coherent state
basis. Identical equations (up to a minus sign in all terms containing $g_k$) 
are obtained for the variables $q_n$ and $h_{k,n}$. 
We have denoted respectively in Eq.~(\ref{pevolution}) and
Eq.~(\ref{fevolution}):
$P_j=-i\frac{\partial E}{\partial p^\star_j}$ and $F^k_j=-i\frac{\partial
E}{\partial f^{\star k}_j}-\frac{i}{2}\left(\frac{\partial E}{\partial
p^\star_j}p^\star_j + \frac{\partial E}{\partial p_j}p_j\right)f^k_j $,
with $E= \la \Psi | H | \Psi \ra$ the average energy, which reads explicitely:
\begin{align}
E = &\f{\Delta}{2} \sum_{n,m} \Big(\conj{p_n} q_m \la f_n | h_m \ra + p_m
\conj{q_n} \la h_n| f_m \ra\Big) \notag \\ \notag
&+ \sum_{n,m} \Big(\conj{p_n}p_m \la f_n | f_m \ra
W_{nm}^f + \conj{q_n} q_m \la h_n | h_m \ra W_{nm}^h \Big) \nn \\
&- \f{1}{2}\sum_{n,m} \Big(\conj{p_n}p_m \la f_n | f_m \ra L_{nm}^f
- \conj{q_n} q_m \la h_n | h_m \ra L_{nm}^h \Big),
\end{align}
where we have defined $W_{nm}^f=\sum_{k>0} \omega_k f_{k,n}^* f_{k,m}$, 
$W_{nm}^h=\sum_{k>0} \omega_k h_{k,n}^* h_{k,m}$, 
$L_{nm}^{f}=\sum_{k>0} g_k (f_{k,n}^* + f_{k,m})$, 
$L_{nm}^{h}=\sum_{k>0} g_k (h_{k,n}^* + h_{k,m})$. 
We now proceed with the implementation of a new and efficient numerical
solution of the dynamical equations.

\subsection{New integration algorithm}
\label{algo}
One can note that the dynamical equation~(\ref{fevolution}) for the displacement
field $f_{k,m}(t)$ is not yet in the proper form where a unique time derivative
$\dot{f}_{k,m}(t)$ is extracted on one side of the set of equations. Achieving such a
decomposition is required for efficient time integration, but considering that
the system under study will require $N_\mr{modes}\simeq1000$ (for accurate
spectral resolution) and $N_\mr{cs}\simeq40$ (for convergence of the quantum
many-body state), a brute force inversion of equations (\ref{pevolution}-\ref{fevolution}) 
would scale prohibitively as $(N_\mr{cs} \times N_\mr{modes})^3$ operations for
each time step. A more efficient algorithm, allowing to cope with a few hundred 
modes was proposed in Ref.~\cite{gheeraert_spontaneous_2016}, used an inversion technique
with only $(N_\mr{cs})^6$ operations, which is favorable provided 
$N_\mathrm{cs}\ll N_\mathrm{modes}$. We present here an improved version of this 
algorithm, which enables us to reach the realistic situation of several thousands 
of modes.

The first step is to multiply Eq.~(\ref{pevolution}) by
$M^{-1}$, with $M$ the overlap matrix $M_{ij} = \braket{f_i|f_j}$:
\begin{eqnarray}
\label{eq:1}
\sum_j M^{-1}_{nj}P_j &=& \dot{p_n}
-\frac{1}{2}\sum_{mj}p_m\kappa_{mj}M^{-1}_{mj}M_{jn}\\
\nonumber 
&=& \dot{p_n}-\frac{1}{2}p_n\sum_q
(\dot{f}_{q,m}f^{\star}_{q,m}+\dot{f}^{\star}_{q,m}f_{q,m})\\
&& + \sum_{mjq}p_m M^{-1}_{nj} M_{jm}f^{\star}_{q,j} \dot{f}_{q,m},
\end{eqnarray}
and similarly for Eq.~(\ref{fevolution}):
\begin{eqnarray}
\nonumber 
\sum_j M^{-1}_{nj}F^k_j &=& p_n \dot{f}_{k,n}+ \dot{p}_nf_{k,n}\\
\nonumber
&&-\frac{1}{2} p_nf_{k,n} 
\sum_q(\dot{f}_{q,n}f^{\star}_{q,n}+\dot{f}^{\star}_{q,n}f_{q,n}) \\
 && +\sum_{jmq}p_m M^{-1}_{nj}M_{jm}f_{k,m}f^{\star}_{q,j} \dot{f}_{q,m}.
\label{eq:2}
\end{eqnarray}
We now substitute Eq.~(\ref{eq:1}) in Eq.~(\ref{eq:2}):
\begin{eqnarray}
\label{eliminate}
p_n\dot{f}_{k,n} &=&
\sum_j(M^{-1}_{nj}F^k_j-f_{k,n} M^{-1}_{nj}P_j) \\
&&-\sum_{mjq}M^{-1}_{nj}M_{jm}p_mf^{\star}_{q,j}\dot{f}_{q,m}\left(f_{k,m}-f_{k,n} \right),
\nonumber
\end{eqnarray}
which allowed to eliminate the complex conjugate time derivative $\dot{f}^\star_{q,m}$.
Eq.~(\ref{eliminate}) is not yet in explicit form since time derivatives of all
possible displacement fields appear in the right hand side.
We define the mode-independent quantities 
$a_{in} = p_n \sum\limits_k f^{\star}_{k,i}\dot{f}_{k,n}$ and
$b_{in} = \sum\limits_k f^{\star}_{k,i}f_{k,n}$ and solve for $a_{in}$ by
inserting Eq.~(\ref{eliminate}) in its expression:
\begin{eqnarray}
a_{in} + \sum_{mj}a_{jm}M^{-1}_{nj}M_{jm}\left(b_{im}-b_{in} \right)&=& A_{in},
\label{eq:ain}
\end{eqnarray}
with $A_{in}=\sum_{jk}f^{\star}_{k,i}\left(M^{-1}_{nj}F^k_j-f_{k,n}M^{-1}_{nj}P_j
\right)$.

After solving the linear system~(\ref{eq:ain}) with the $(N_\mr{cs})^2$ unknown 
parameters $a_{in}$, the evolution equation for each displacement field is then 
cast into explicit form: 
\begin{eqnarray}
\nonumber
p_n\dot{f}_{k,n} &=&
\sum_{jk}(M^{-1}_{nj}F^k_j-f_{k,n}M^{-1}_{nj}P_j) \\
&&-\sum_{mj}M^{-1}_{nj}M_{jm}a_{jm}(f_{k,m}-f_{k,n}),
\label{eq:fdot} 
\end{eqnarray}
which can be integrated numerically using an RK4 method.
The numerical inversion of the system~(\ref{eq:ain}) can be sped up below
the naive $(N_\mr{cs})^6$ cost by defining
$d_{in}=\sum_j M^{-1}_{nj}M_{jm}a_{jm}$ and 
$\alpha_{inm}=\sum_lM^{-1}_{in}M_{ln}(b_{lm}-b_{ln})$, so that we can solve
a linear system for $d_{in}$:
\begin{equation}
    \sum_{mj}\left(\delta_{ij}\delta_{nm}+\alpha_{inm}\delta_{nj}
\right)d_{mj}=
\sum_j M^{-1}_{nj}M_{jm}A_{jm}
\end{equation}
which assumes a sparse form suitable for Krylov based methods (provided a
good preconditioner can be found).

\subsection{Incoming and outgoing many-body states}
\label{AppState}
Combining the incoming coherent state, described by the displacement $z_k$
in Eq.~(\ref{zinput}), with the static polarization cloud wavefunction Eq.~(\ref{Psi_GS}) 
can be done by transforming the incoming signal in the even/odd basis (see
Sec.~\ref{formalism}). For the spin-up projection of the wavefunction, we
readily find: 
\eq{
 \ket{\Psi_\ua} &= D(z^{e}) D(z^{o}) \ket{\Psi^{\mrm{GS}}_\ua} \\
&= e^{\sum_{k>0}z_{k}^{e}a_{k}^{e\dagger}-c.c.} \sum_m^{{N^{\mrm{GS}}_{\textrm{cs}} } } 
p^\mrm{GS}_m e^{\sum_{k>0}f_{k,m}^\mr{GS}a_{k}^{e\dagger}-c.c.} \ket{0}_e\ket{z^o}_o
\notag 
}
which can be recombined using the standard relation 
$e^{A}e^{B}=e^{A+B}e^{\frac{1}{2}[A,B]}$, valid as the commutator
here is only a number.
The initial state associated to the $\ua$ qubit state thus reads: 
\begin{eqnarray}
 \ket{\Psi_\ua}
&=& \sum_m^{{N^{\mrm{GS}}_{\textrm{cs}} } } 
p^\mrm{GS}_m e^{\frac{1}{2}\sum_{k>0}
\left(z_{k}^{e}f_{k,m}^{\mr{GS}*}-z_{k}^{e*}f_{k,m}^\mr{GS}\right)}\\
\nonumber
&& \times \, e^{\sum_{k>0}\left[\left(f_{k,m}^\mr{GS}+z_{k}^{e}\right)
a_{k}^{e\dagger}-\left(f_{k,m}^\mr{GS}+z_{k}^{e}\right)^{*}a_{k}^{e}\right]}\ket{0}_e\ket{z^o}_o .
\end{eqnarray}
For the spin-down projection, one simply replaces $f_{k,m}^\mr{GS}$
by $-f_{k,m}^\mr{GS}$ without changing the sign of $z_{k}^{e}$, 
so that our total initial wavefunction is given by Eq.~(\ref{PsiInit}).

The outgoing wavepacket is constructed in a similar spirit:
\begin{equation}
\label{PsiOUT}
\ket{\Psi^{\mrm{OUT}}} = 
\sum_{m=1}^{{N^{\mrm{OUT}}_{\textrm{cs}} } } 
p^\mrm{OUT}_m e^{\frac{1}{2}\sum_{x>0}
\left(f_{x,m}^{\mr{OUT}}a_x^\dagger-f_{x,m}^{\mr{OUT}*}a_x\right)}\ket{0},
\end{equation}
where we have written the displacements of the outgoing state in real space 
because they have no spatial overlap with the real space modes that
populate the many-body ground state (working in momentum space would complicate
the analysis). This decoupling occurs in fact when the wavepacket reaches distances 
away from the qubit that are larger than the inverse Kondo energy~\cite{snyman_robust_2015}, 
or said otherwise, that are larger than the entanglement cloud around the qubit.
Clearly the quantum many-body character of the scattering process is encoded in 
the sum over more than a unique coherent state, in contrast to the incoming 
wavepacket~(\ref{WaveIn}) that is characterized by a single coherent state (namely 
a classical-like signal). Contrarily to the driven dynamics for an isolated
few level quantum system, this long time equilibration between the many-body ground 
state and the wavepacket is physically expected because the waveguide acts as a bath 
for the dressed two-level system, and thus provides a natural pathway for
relaxation, even in a many-body system.

Extracting the wavepacket contribution~(\ref{PsiOUT}) from the long-time
wavefunction~(\ref{PsiFinalTime}) can be performed as follows. The complete set
of displacements $\{f_{k,n}^{e}(T),h_{k,n}^{e}(T)\}$ in the even sector at a
fixed long time $T$ for the full wavefunction~(\ref{Psi}) are first Fourier
transformed to real space using (\ref{FT}). The local photon density $n(x)$
associated to these displacements is sketched in Fig.~\ref{waveguide}: photons
are either bound statically near the qubit (associated to the dressed vacuum) or 
travel in the outgoing wavepackets. The displacements are then simply set to zero 
in the region surrounding the qubit, and Fourier transformed back to the momentum basis. 
Due to factorization~(\ref{PsiFinalTime}), the outgoing wavefunction is recovered, up 
to a normalization factor, which is supplemented accordingly. The even 
modes thus obtained and the trivial odd mode wavefunctions are finally combined together in the case 
of the incoming wavepacket, allowing reconstruction of the full outgoing wavefunction 
for the physical waveguide. 

\subsection{Convergence properties}
\label{AppConvergence}

\begin{figure}[t]
\includegraphics[width=1.0\linewidth]{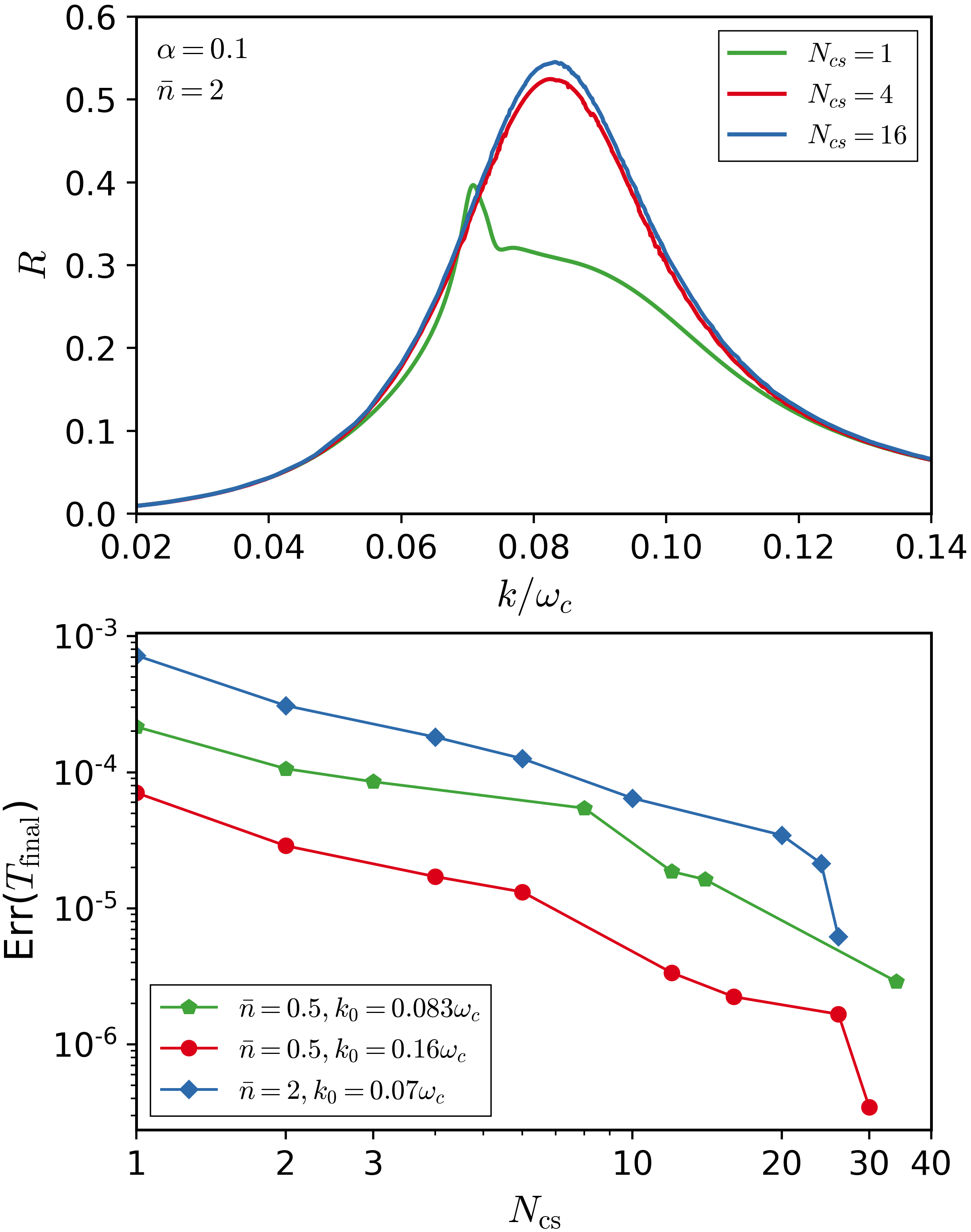}
\caption{Top panel: Power reflection spectrum shown for different number of
coherent states $N_\mr{cs}=1,4,16$ included in the MCS wavefunction~(\ref{Psi}), with
the same parameters as in Fig.~\ref{saturation} in the case of $\bar{n}=2$
photons in the incoming beam. Bottom panel: Convergence of the error defined
in the text at the final time $T_{\rm{final}}$ of the simulations, as a function
of coherent state number in the wavefunction for various incoming momenta and
power.}
\label{convergence}
\end{figure}

Assessing the good convergence of the numerical results is important to
gain confidence in the time-dependent variational MCS technique Indeed, we find that using too few variational 
parameters imposes strong constraints on the dynamics, which may result 
in unphysical behavior and numerical artifacts.
One delicate test is the strong power saturation spectrum shown in 
Fig.~\ref{saturation} of the main text. Indeed, the calculations that use only 
a single coherent state, as done in a previous publication~\cite{bera_dynamics_2016}, 
are found to be problematic in the strong power regime. 
This behavior is illustrated in the top panel of Fig.~\ref{convergence}, showing 
the power reflection spectrum as a function of incoming frequency at a strong input power 
($\bar{n}=2$) for three different values of 
the number of coherent states $N_\mr{cs}=1,4,16$. 
The computation with $N_\mr{cs}=1$ is indeed quite noisy and imprecise, and a 
smooth and converged curve is only obtained at $N_\mr{cs}=16$. We find that
the inelastic spectra shown in Fig.~\ref{off_res_Fock} are also delicate to
compute, because they consist of a tiny fraction of the total signal, and 
encode complex quantum states. A relatively large number of coherent state is
also necessary here for success, even at small input power.

An unbiased criterion for the convergence of our algorithm for this
non-equilibrium many-body dynamics is also shown in the lower panel of
Fig.~\ref{convergence}. Here we demonstrate that the error with respect
to the exact Schr\"odinger dynamics vanishes with the number of coherent states. 
The error is defined~\cite{gheeraert_spontaneous_2016} by the squared norm
\eq{
\mathrm{Err}(t)\equiv \big<\Phi(t)|\Phi(t)\big>
\label{error}
}
of the auxiliary state $|\Phi(t)\big>\equiv(i\partial_t -H)|\Psi(t)\big>$.
Indeed, this error decreases steadily and scales as
$[N_{\textrm{cs}}]^{-2}$. For the off-resonant case of Fig.~\ref{off_res_Fock}
(see bottom curve in the lower panel of Fig.~\ref{convergence}) we managed to reach
an error of the order of $10^{-7}$. 

\subsection{Protocol for adding coherent states during the time evolution}
\label{AppProtocol}

\begin{figure}[t]
\includegraphics[width=1.0\linewidth]{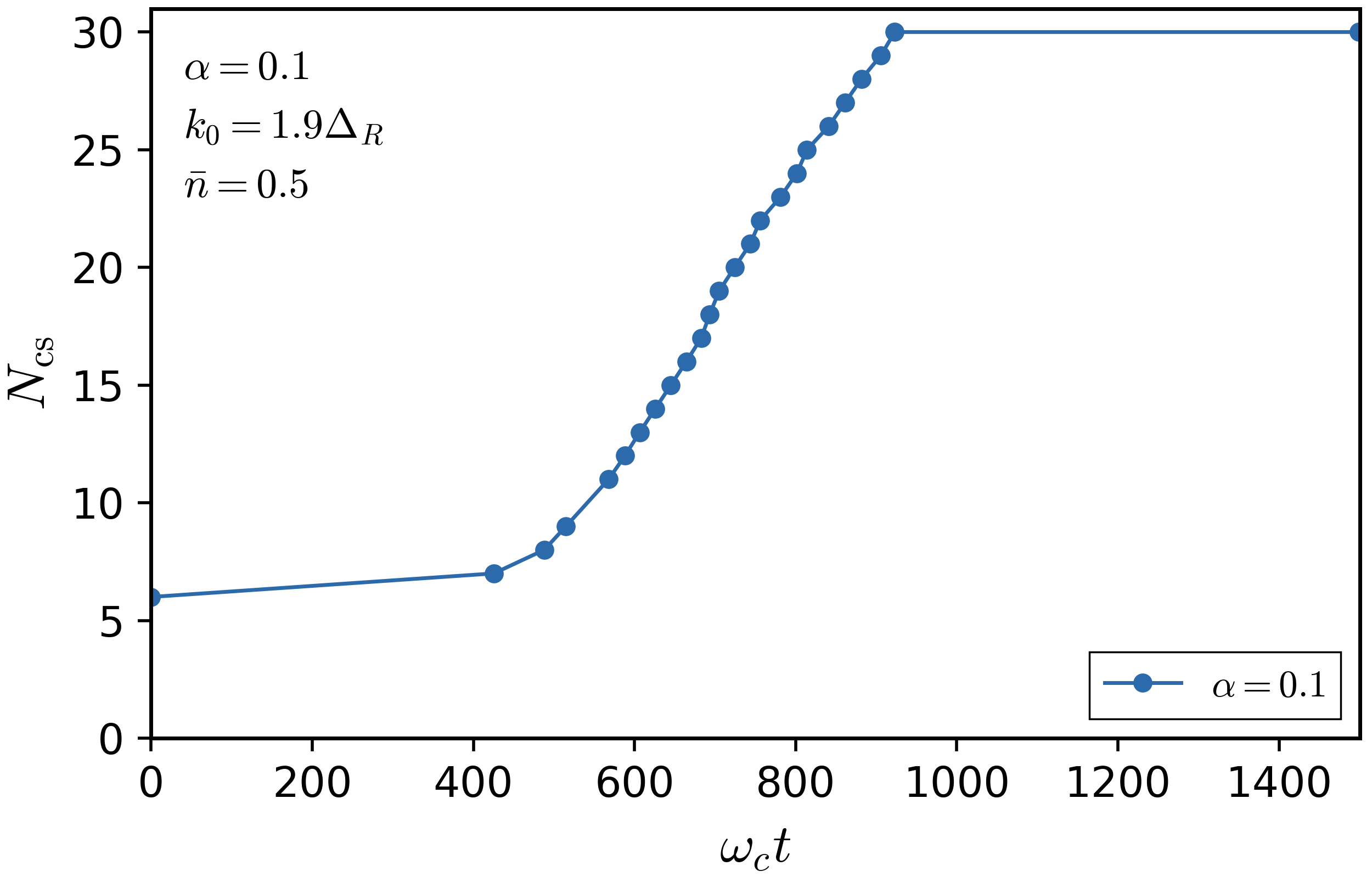}
\caption{Number of used coherent states in the MCS wave-function defined in Eq.~(\ref{Psi}) 
as a function of time, for the off-resonant $k_0=0.16\w_c$ simulation. The initial 6 coherent 
states present are those required for describing the dressed ground state of the 
spin-boson model as well as the incoming wave-packet. Subsequent additions of coherent 
states start only as the wave-packet starts to impinge on the qubit. The addition process 
stops as the wavepackets exit the interaction region.}
\label{ncs}
\end{figure}

Because the coherent state basis is over-complete, all the coherent
states required for good convergence (typically $N_\mr{cs}>16$) cannot be initialised 
simultaneously at the initial time. Indeed, two coherent states with identical displacements 
will result in a singularity in the matrices to be inverted for solving the dynamics,
due to a vanishing determinant. 
During the initial stage of the dynamics, this is not an issue, as only a small 
number of coherent states (typically 6 to 10) is needed to describe the static 
many-body cloud and the incoming coherent state. After some time however, the 
wavepacket starts to interact with the dressed qubit, which would increase the 
error should the number of coherent states remain the same.
Therefore, to account for the emerging complexity of the many-body scattered
state, we progressively increase the number of coherent states $N_{\textrm{cs}}$
in the MCS state vector (Eq.~(\ref{Psi})), initializing the newly added coherent
states in a bosonic vacuum configuration with zero weight. Thus, the addition of
a new set of variational displacement does not immediately affect the dynamics,
but provides the necessary freedom to our variational algorithm for
maintaining a minimal error at later times. This procedure is illustrated in
Fig.~\ref{ncs}.

The right time for a coherent state to be added is found by monitoring the
error, which was defined by Eq.~(\ref{error}), and by defining an error increment
$\mr{Err}_\mr{max}$ at which the coherent state should be added. Whenever
$\mr{Err}(t)-\mr{Err}_{\mr{ref}}>\textrm{Err}_\mr{max}$ during the time
evolution, where Err$_{\mr{ref}}$ is the error right after the previous coherent
state was added, one simply adds a coherent state with displacement
$f_m^k=h_m^k=0$ 
and weights $p_m=q_m \simeq 10^{-6}$ in Eq.~(\ref{Psi}). The near-zero amplitude ensures that 
this new coherent state only changes the wave-function negligibly at the time 
it is added. Empirically, we find the value of $\mr{Err}_\mr{max}=10^{-7}$ to be adequate. 
The system, through the variational principle, will subsequently have the possibility 
to increase the displacements and weights according to the requirements of the quantum 
trajectory. As an example, a plot of the number of coherent states as a function of 
time for the off-resonant $k_0=0.16\w_c$ simulation in Fig. \ref{off_res_Fock} is given 
in Fig. \ref{ncs}.

\subsection{Calculation of Number Resolved Spectra}
\label{AppNumresolve}

To assess the nature of particle production in the scattering process, we
analyze the inelastic spectrum in terms of Fock states $\ket{N}$.  First,
consider the general expansion of the multi-mode outgoing
wavefunction~(\ref{PsiOUTk}) in terms of number states:
\begin{eqnarray}
\ket{\Psi^\mr{OUT}} & = &\gamma \ket{0} + \sum_k \alpha_k a_k^\dag \ket{0} 
+ \sum_{k_1,k_2} \alpha_{k_1,k_2} a_{k_1}^\dag a_{k_2}^\dag \ket{0} \notag \\ 
&& + \sum_{k_1,k_2,k_3} \alpha_{k_1,k_2,k_3} a_{k_1}^\dag a_{k_2}^\dag a_{k_3}^\dag \ket{0} 
+\ \ \mrm{...}
\end{eqnarray}
It can then easily be verified that the 1-photon amplitude is given by:
\begin{equation}
\alpha_k = \bra{0} a_k \ket{\Psi^\mr{OUT}} = \sum_n p_n f_{k,n} \braket{0 | f_n},
\label{alpha1photon}
\end{equation}
and that the scattering amplitude for a generic $N$-photon state is:
\begin{equation}
\label{fock_amplitude}
\alpha_{k_1,...,k_N} = \frac {1}{N!} \bra{0} a_{k_1} ...\ a_{k_N} \ket{\Psi^\mr{OUT}},
\end{equation}
which can be obtained straightforwardly from the algebraic identities of coherent states. 
For the sake of clarity, we have dropped the OUT labels on $p_n$ and $f_{n,k}$.
From the multi-photon amplitudes, we can then compute the probability 
distribution for finding a photon in a given $k$ mode, according to the
various Fock contents of the total wavefunction:
\begin{eqnarray}
\braket{n_k}_\mr{1photon} &=& |\alpha_{k}|^2 \notag, \\
\braket{n_k}_\mr{2photon} &=& 4 \sum_{k1} |\alpha_{k,k1}|^2 \notag, \\
\braket{n_k}_\mr{3photon} &=& 18 \sum_{k2,k3} |\alpha_{k,k2,k3}|^2.
\end{eqnarray}
These Fock resolved inelastic contributions
$\braket{n_k}_\mr{Nphoton}$, with $N=2,3,4$ are displayed as full lines in 
Fig.~\ref{off_res_Fock} (note that the outgoing $N=1$ process is purely elastic
and is not shown).

\subsection{Calculation of $g_2(\tau)$}
\label{App-g2calc}

In this appendix, we give some details on the calculation of the correlation
function $g_2(\tau)$ when using the MCS approach. First, since we take the speed
of light $c=1$, $\tau$ is just the distance traveled by radiation in time
$\tau$. Inserting the MCS expansion Eq.~(\ref{Psi}) into definition (\ref{g2}),
we obtain a compact expression for the autocorrelation function in terms of the
real space displacements $f_n^x$:
\begin{equation}
g_2(\tau) = \frac{\sum_{m,n} p_n^* p_m (f^{x}_n)^* 
(f^{x+\tau}_n)^* f^{x+\tau}_m f^x_m \braket{f_n |
f_m}}{\braket{n(x)}\braket{n(x+\tau)}},
\end{equation}
with the local photon number
\begin{equation}
\braket{n(x)}=\braket{a^\dag_x a_x}=\sum_{m,n} p_n^* p_m (f^{x}_n)^* f^{x}_m
\braket{f_n|f_m}.
\end{equation}
In the simulations performed to compute this quantity we used a sharp cutoff
$\Theta(\omega_c-\omega)$ for the dispersion relation instead of the exponential
cutoff $\rm{e}^{-\omega/\omega_c}$ which we 
defined in Eq.~(\ref{gk}). Note that using the hard cut-off results in a
slightly lower value of the renormalised qubit energy $\Delta_R$, than with the
exponential cutoff. This allowed us to decrease the numerical cost and
therefore attain a higher number of coherent states, $N_{\rm{cs}}=40$, which was 
necessary because second-order 
correlations are more challenging to converge than average photon numbers. The
simulations were stopped at a timescale $T=1250/\w_c$ long enough that the 
wavepacket is located far away from the dressed qubit, and we chose the spacial
point $x=-681$ in Eq. (\ref{g2}), so as to  keep the range of the function near
the center of the wavepacket.  We finally note that spurious effects associated
with the finite spatial extension of the 
wavepacket (due to $\sigma\neq0$) lead to the small oscillations seen in
Fig.~\ref{anti} at longer times.

\section{Further analysis of non-linear emission}

\subsection{Detailed off-resonant conversion spectra}
\label{GoldsteinApp}

\begin{figure}[htb]
\label{off_res_comp}
\includegraphics[width=1.0\linewidth]{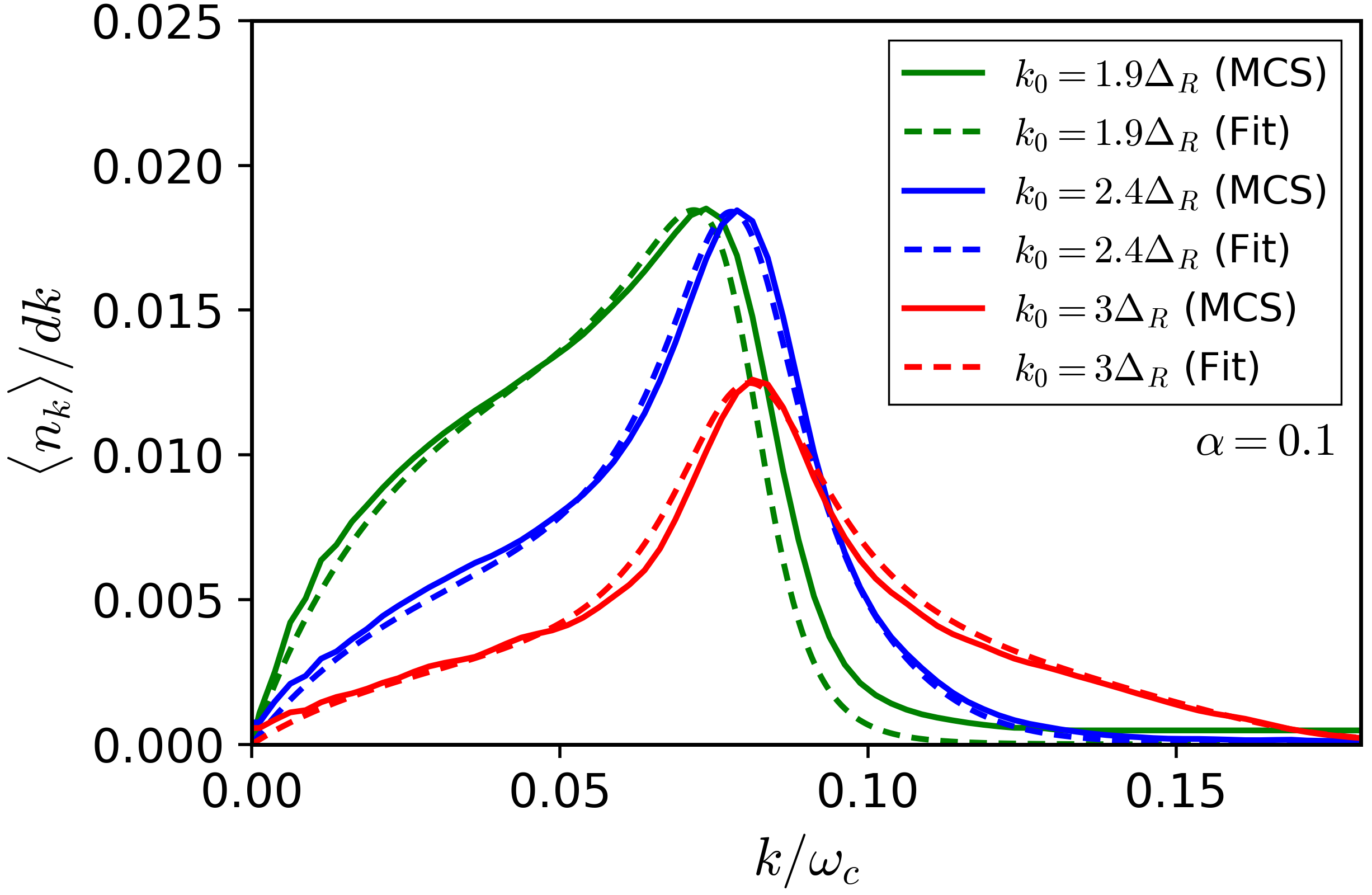}
\caption{Low-energy inelastic spectrum for several off-resonant values of 
the incoming momentum $k_0=0.16,0.20,0.25\w_c$ obtained using the MCS technique
(with the same parameters as in Fig.~\ref{off_res_Fock}), together with a
comparison to the analytical formula~(\ref{CompGoldstein}), using a fitted and
momentum-dependent linewidth $\gamma_R(k_0)$. }
\end{figure}
We proceed here with a systematic study of particle production spectra
in the off-resonant case, as a function of incoming momentum $k_0$ 
(see Fig.~\ref{off_res_comp}).
A weak coupling calculation of the one photon to three photon conversion process
(see Fig.~\ref{diagrams}) was given in the $\alpha\to0$ limit in 
Ref.~\cite{goldstein_inelastic_2013}. We have found that this theory can
quantitatively account for our data at small $\alpha$  upon two important modifications. 
First, as already seen by the frequency shift in the reflection spectrum in
Fig.~\ref{saturation}, one must replace the bare qubit frequency $\Delta$ by 
the renormalized quantity $\Delta_R$ within the analytical results given 
by the perturbative approach.
Second, the golden rule value for the qubit linewidth appearing in the transmission
lineshape, given by $\Gamma=\pi \alpha \Delta$ at small $\alpha$, cannot be
used.
For the elastic response, one can use reliably $\Gamma_R=\pi \alpha 
\Delta_R$ up to moderate values of $\alpha$. However, we find that the renormalized 
broadening parameter $\gamma_R^\mr{inel}$ entering the inelastic response function 
for fixed value of the incoming momentum $k_0$ is not given by $\Gamma_R$, but
rather displays a strong momentum dependence, $\gamma_R^\mr{inel} = \gamma_R(k_0)$. 
This is not completely unexpected, since a consistent calculation should include 
the full momentum variation of the self-energy, and we found that the theory of 
Ref.~\cite{goldstein_inelastic_2013} is very sensitive to the way the inelastic 
regularization is implemented.
For the present purpose, we will only use a phenomelogical model that uses (as
fitting parameters) only two renormalized quantities $\Delta_R$ and $\gamma_R(k_0)$ 
within the perturbative formula:
\begin{eqnarray}
\label{CompGoldstein}
\braket{n_k}_\mr{3photon} &=& \frac{\alpha^4}{8} \Delta_R^2 
\int_0^{k_0-k}\!\!\!\!\! \mr{d}k_1 \;
k k_0 k_1 k_2 \\
\nonumber
&& \hspace{-2cm} \Big| 
\frac{k_0 k k_1 k_2 - k_\Delta^2 (k^2+k_1^2+k_2^2+kk_1+kk_2+k_1k_2)+3k_\Delta^4}
{(k^2-k_\Delta^2) (k_0^2-k_\Delta^2) (k_1^2-k_\Delta^2)
(k_2^2-k_\Delta^2)}\Big|^2,
\end{eqnarray}
with $k_2=k_0-k_1-k$ and $k_\Delta=\Delta_R+i\gamma_R(k_0)$, with $\Delta_R$ the
renormalized qubit frequency and $\gamma_R(k_0)$ the linewidth describing
the inelastic spectrum, which is fitted from our numerical data.
The resulting comparison is shown in Fig.~\ref{off_res_comp}, with excellent quantitative 
agreement.

\subsection{Detailed on-resonant conversion spectra}

\begin{figure}[b]
\includegraphics[width=1.0\linewidth]{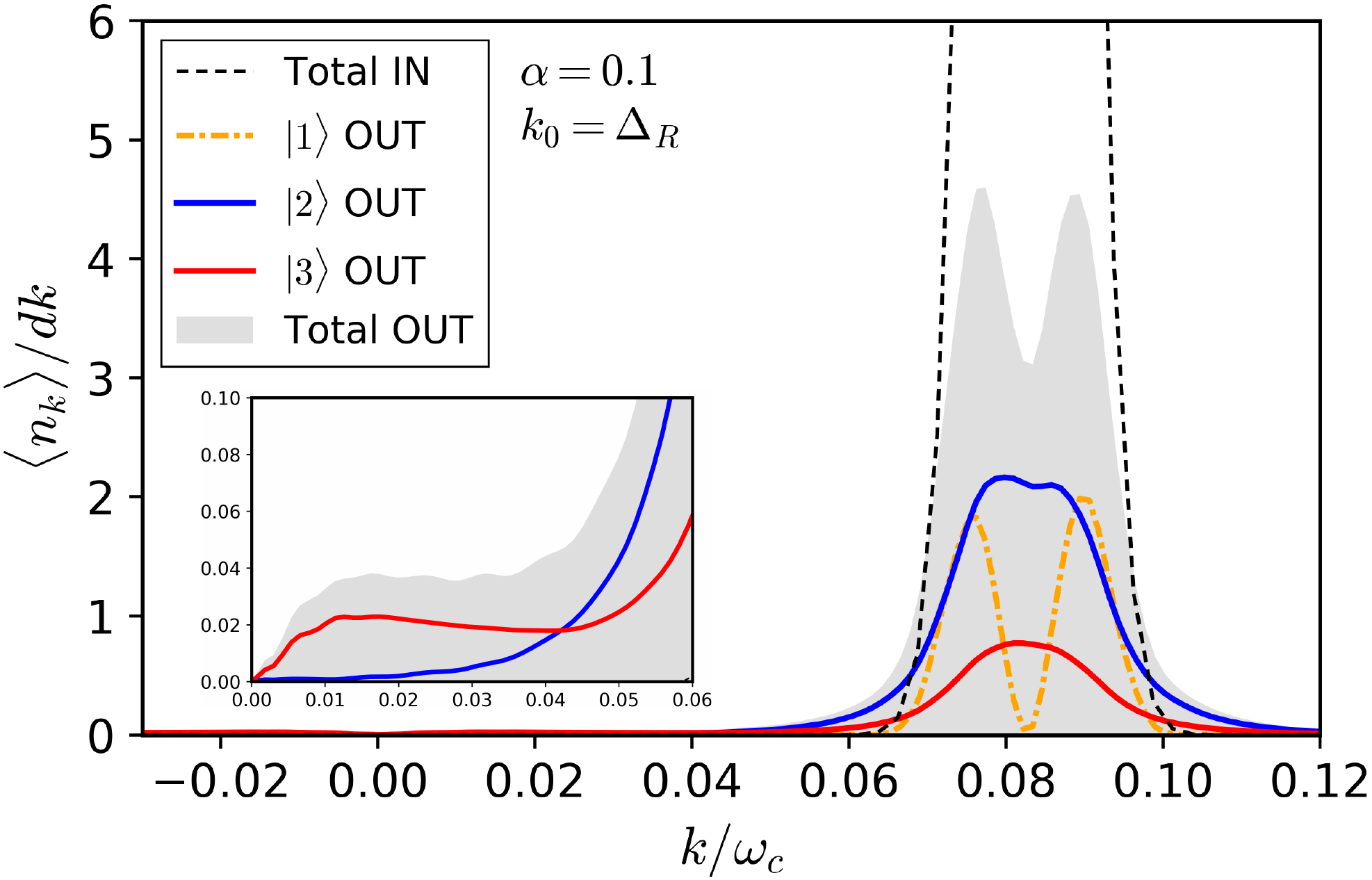}
\caption{On-resonance transmission spectrum for an incoming photon energy that matches 
the renormalized qubit excitation, $k_0=0.08\w_c=\Delta_R$. The 
simulation required 2500 modes with the use of 34 coherent states (all other 
parameters are identical to the ones in Fig.~\ref{off_res_Fock}).
The black dashed curve corresponds to the nearly monochromatic incoming wave-packet.
Because of the wavepacket finite linewidth $\sigma$, a small fraction of
one-photon states is still transmitted (dot-dashed line), despite being
on resonance. The two-photon contribution (top full line) presents wider inelastic
wings, that extend beyond the width $\sigma$, and that are parametrically larger
in amplitude than the off-resonant signal of Fig.~\ref{off_res_Fock}. 
The 3-photon continuum is magnified in the inset.}
\label{on_res}
\end{figure}

We consider here the detailed photonic content of the emission spectra in the resonant 
case where the incoming photon energy
$k_0=\Delta_R$ matches the renormalized atomic transition energy.
Fig.~\ref{on_res} shows the total transmitted signal as well as its
decomposition in terms of number states with $N=1,2,3$ photons.
%We first notice that as expected that {\it exactly} on resonance, at $k=k_0$,
%the transmission of the one photon state is exactly zero, while the part of the
%signal that is slighly off resonant (due to the finite spectral width of our
%incoming pulse), gets only partially transmitted. 
Not surprisingly, the two-photon amplitude in this regime is strongly enhanced with respect
to the off-resonant situation of Fig.~\ref{off_res_Fock}. One-photon
contributions are also observed as two side-bands away from the resonance $\Delta_R$, 
which are due to the finite width of the incoming wave-packet. The resonant 1-photon states 
(at exactly $k=k_0$) are completely reflected, as expected.
%The on-resonant regime is a promising one to observe experimentally non-linear 
%effects, because of the enhanced magnitude 
%of the inelastic signal, as compared to the weaker off-resonant situation of 
%Fig.~\ref{off_res_Fock}. 
In the resonant case, $2\to2$ RWA frequency conversion gives rise 
to the broader wings (extending clearly beyond the linewidth $\sigma$ of the pump), 
as seen in the $N=2$ curve of Fig.~\ref{on_res}.
As in the off-resonant case, the resonant scattered spectrum also presents a
3-photon low energy continuum, as can be seen from the inset. The shape however
does not present any sharply peaked feature, since this time the continuum does
not contain the resonant frequency $k=\Delta_R$ at which the qubit spontaneously
reemits. Instead, the spectrum is more flat, implying the single photon splits
more uniformly into all the possible $(k_1,k_2,k_3)$ allowed by the $1\to 3$ process
of Fig.~\ref{diagrams}. Interestingly, the magnitude of this 3-photon continuum
is of the same order of magnitude as in the off-resonant case of
Fig.~\ref{off_res_Fock}, since non-linear processes are here intensified by
having an on-resonant input, which compensates for the absence of an enhancing
resonant frequency in the output below $k_0$. Again, this particle production
process dominates the RWA contribution, here only away from the probe frequency. 

%%%%%%%%%%%%%%%%%%%%%%%%%%%%%%%%%%%%%%%%%%%%%%%%%%%%%%%%%%


\end{document}